\documentclass[twocolumn]{aa}

\usepackage{graphicx}
\usepackage{txfonts}

\begin{document}

\title{Modeling optical and UV polarization of AGNs I.\\
       Imprints of individual scattering regions}

\author{Ren\'e W. Goosmann \inst{1,2} \and C. Martin Gaskell \inst{3}}

\authorrunning{Goosmann \& Gaskell}

\titlerunning{Modeling AGN Polarization}

\offprints{Ren\'e W. Goosmann}

\institute{
    Astronomical Institute of the Academy of Sciences, Bo{\v c}ni II 1401,
    14131 Prague, Czech Republic\\
    \email{goosmann@astro.cas.cz}
   \and
    Observatoire de Paris - Meudon, 5 place Jules Janssen,
    92190 Meudon, France
   \and
    Department of Physics \&  Astronomy, University of Nebraska,
    Lincoln, NE 68588-0111, USA\\
    \email{mgaskell1@unl.edu}
   }

\date{Received June 2005; accepted December 2006}

\abstract{Spectropolarimetry of AGNs is a powerful tool for studying
the structure and kinematics of the inner regions of quasars.}{We wish
to investigate the effects of various AGN scattering region geometries
on the polarized flux.}{We introduce a new, publicly available Monte
Carlo radiative transfer code, {\sc Stokes}, which models polarization
induced by scattering off free electrons and dust grains. We model a
variety of regions in AGNs.}{We find that the shape of the funnel of
the dusty torus has a significant impact on the polarization
efficiency. A compact torus with a steep inner surface scatters more
light toward type-2 viewing angles than a large torus of the same
half-opening angle, $\theta_0$. For $\theta_0 < 53\degr$, the
scattered light is polarized perpendicularly to the symmetry axis,
whilst for $\theta_0 > 60\degr$ it is polarized parallel to the
symmetry axis. In between these intervals the orientation of the
polarization depends on the viewing angle. The degree of polarization
ranges between 0\% and 20\% and is wavelength independent for a large
range of $\theta_0$. Observed wavelength-independent optical and
near-UV polarization thus does not necessarily imply electron
scattering. Spectropolarimetry at rest-frame wavelengths less than
2500~\AA~may distinguish between dust and electron scattering
but is not conclusive in all cases.  For polar dust, scattering
spectra are reddened for type-1 viewing angles, and made bluer for
type-2 viewing angles. Polar electron-scattering cones are very
efficient polarizers at type-2 viewing angles, whilst the polarized
flux of the torus is weak.}{We predict that the net polarization of
Seyfert-2 galaxies decreases with luminosity, and conclude that the
degree of polarization should be correlated with the relative strength
of the thermal IR flux.  We find that a flattened, equatorial,
electron-scattering disk, of relatively low optical depth, reproduces
type-1 polarization. This is insensitive to the exact geometry, but
the observed polarization requires a limited range of optical depth.}


\maketitle

\section{Introduction}

One of the foremost problems in AGN research is that the innermost
regions of AGNs cannot be resolved in the optical and UV with
current technology. However, the light of AGNs is polarized over a
broad wavelength range, and this allows us to put important
constraints on the geometry of the emitting and scattering regions.
Spectropolarimetric observations giving the detailed wavelength
dependence of the polarized flux give further clues to the nature of
the polarizing mechanism.

Our inferences of the innermost structures of AGNs have so far been
obtained indirectly.  Rowan-Robinson (\cite{rowan-robinson77})
suggested that AGNs are surrounded by a dusty torus and in the same
paper he gives a suggestion by M. V. Penston that Seyfert~2 galaxies
are seen close to edge-on so that the active nucleus is obscured by
the torus. Support for this picture came from the important
discovery by Keel (\cite{keel80}) that Seyfert~1 galaxies (active
galaxies showing a broad-line region; BLR) are preferentially seen
face-on. Keel (\cite{keel80}) also investigated absorption effects
inside the host galaxies and emphasized the need of additional
nuclear absorption in Seyfert galaxies with respect to normal
spirals. Keel's work led to further confirmation of the importance
of orientation effects (Lawrence \& Elvis \cite{lawrence82}; De
Zotti \& Gaskell \cite{dezotti85}). Since then, the dusty-torus
model has become the standard unified model (see Antonucci
\cite{antonucci93}) dividing AGNs into two sub-types: ``type-1''
AGNs which are seen close to face-on, and ``type-2'' AGNs which are
seen close to edge-on. In type-1 AGNs the central energy source and
its surroundings (e.g., the BLR) can be seen, whilst in type-2 AGNs
the torus blocks our direct view of these inner regions. While this
obscuration and the IR emission from the torus are the most obvious
effects of the torus, scattering from the dust will add polarized
flux. The polarization spectrum of an optically thick dusty torus
has been the subject of several modeling projects (Kartje
\cite{kartje95}, Wolf \& Henning \cite{wolf99}, Watanabe et al.
\cite{watanabe03}).

When Dibai \& Shakhovskoy (1966) and Walker (1966) discovered
optical polarization of AGNs, it was initially taken to be evidence
of optical synchrotron emission, since synchrotron radiation has a
high intrinsic polarization.  However, Angel et al. (\cite{angel76})
found the Balmer lines in NGC 1068 to be polarized similarly to the
continuum, thus implying that scattering was responsible for the
polarization of both the lines and continuum.  The difference they
found in polarization between the narrow-line region (NLR) and BLR
places the scattering region outside the BLR, but inside the NLR

When light is scattered, the angle of polarization depends on the
direction of the last scattering, so one expects the angle of
polarization to be related to the structure of the AGN. Stockman,
Angel, \& Miley (\cite{stockman79}) made the seminal discovery that
for low-polarization, high optical luminosity, radio-loud AGNs, the
optical polarization position angles tend to align {\it parallel} to
the large-scale radio structure. Although they interpreted this as a
consequence of optical synchrotron emission, they also suggested
that polarization from an optically-thin, non-spherically-symmetric
scattering region near the source of optical radiation was another
possibility.

Antonucci (\cite{antonucci82}) pointed out that whilst many radio
galaxies showed a similar parallel alignment of the polarization and
radio axes, there was, unexpectedly, a population showing a {\it
perpendicular} relationship. It was subsequently shown (Antonucci
\cite{antonucci83}) that relatively-radio-quiet Seyfert galaxies
show a similar dichotomy between the predominantly, but not
exclusively, parallel polarization in face-on type-1 Seyferts and
the perpendicular polarization of type-2 Seyferts (see Antonucci
\cite{antonucci93}, \cite{antonucci02} for reviews). These
discoveries made a synchrotron origin of the polarization much less
likely.

Polarization perpendicular to the axis of symmetry is easily produced
by scattering off material close to the axis. There is good
observational evidence for the existence of ionization cones along the
polar axis in numerous objects (see Kinney et al.  \cite{kinney91} and
reference therein). Polar scattering has been particularly well
studied in the Seyfert-1 galaxy NGC~1068.  Antonucci \& Miller
(\cite{antonucci85}) made the key discovery that the polarized-flux
spectrum can offer a periscope view of type-2 AGNs because much of the
polarized flux originates inside the torus.  Detailed HST polarimetry
has revealed the polarization structure of the ionization cones (see
Capetti et al. 1995a,b; Kishimoto \cite{kishimoto99}).

The detection of a hidden broad-line region in NGC~1068 by Antonucci
\& Miller (\cite{antonucci85}) was of great importance for AGN
research since it provided strong support for the unified theories
of AGN activity. More hidden type-1 nuclei have subsequently been
found by analysis of their polarized-flux spectra (see e.g., Miller
\& Goodrich, 1990; Tran, Miller, \& Kay, 1992; Hines \& Wills, 1993;
Kay, 1994; Heisler, Lumsden, \& Bailey, 1997; Tran, 2001; Smith et
al., 2004). Similar work on the radio galaxy 3C 321 was done by
Young et al. (\cite{young96b}), and Tran et al. (\cite{tran99})
could identify an active nucleus inside an ultra-luminous infra-red
galaxy using spectropolarimetry. Recently, hidden type-1 nuclei have
also been  found in five type-2 quasar candidates (Zakamska et al.
\cite{zakamska2005}).

The new generation of large telescopes is delivering
spectropolarimetry of emission line profiles with good velocity
resolution (see, for example, the spectropolarimetry of the Seyfert
1.5 galaxy NGC 4151 presented by Martel \cite{martel98}, and the
atlas of spectropolarimetry of Seyfert galaxies presented by Smith
et al. \cite{smith02}). Examination of velocity-dependent
polarization of emission lines promises to reveal valuable
information about the geometry of the BLR (Smith et al.
\cite{smith05}). Similarly, spectropolarimetry of quasar absorption
lines helps constrain the geometry of broad absorption line QSOs
(Goodrich \& Miller 1995, Cohen et al. 1995, Hines \& Wills 1995,
Ogle et al. 1999).

In order to understand these many facets of AGN polarization, and
their implications for the underlying geometry, theoretical modeling
is necessary. Analytical approaches to radiative transfer that have
been carried out so far are generally limited to the consideration
of single-scattering models. Computer simulations are needed to
investigate multiple-scatterings. In this paper we describe {\sc
Stokes}, a new general-purpose,
publicly-available,\footnote{http://www.stokes-program.info/} Monte Carlo
code for modeling wavelength-dependent polarization in a
wide variety of scenarios, and we present some results of our study
of AGN polarization.

In this paper we confine ourselves to using {\sc Stokes} for
calculating the polarization imprints of basic constituents of the
unified scheme. We compute the polarization spectrum of dusty tori
with various geometries and opening angles, and we consider
scattering in polar cones and electron disks. We investigate the
effects of geometrical shape and optical depth of given regions.
None of our models are intended to reproduce observational
polarimetric data for any specific object. Rather, we want to
investigate general constraints on the scattering regions and the
geometry of AGNs. We discuss consequences for the observed
polarization dichotomy between type-1 and type-2 objects. We leave
aside the question of interactions between different types of
scattering regions for paper II (Goosmann \& Gaskell, in
preparation) where we also conduct more detailed modeling of AGNs in
the unified scheme.

The present paper is organized as follows: in section 2 we summarize
modeling of optical and UV polarization of AGN and the main results
obtained previously. Section 3 describes our code {\sc Stokes}. In
section 4 we present modeling results for equatorial, toroidal dust
distributions. Section 5 is dedicated to electron and dust
scattering in polar double-cones. In section 6 we investigate the
polarization signature of equatorial regions for electron
scattering. Our results are discussed in section 7 and we give some
conclusions in section 8.

\section{Previous codes and modeling}
\label{sec:review}

In this section, we briefly summarize some recent AGN polarization
modeling codes which we will compare our {\sc Stokes} modeling with.

Young et al. (\cite{young95}, \cite{young96a}), Packham (1997), and
Young (\cite{young00}) developed an analytical radiative transfer
model, the Generic Scattering Model (GSM), for polarization modeling.
The model is based on the unified AGN model. Extended emission
regions can be defined, and scattering processes as well as dichroic
absorption are considered. The modeled geometries include toroidal,
disk-like, and conical regions of dust and free electrons. For
scattering material in motion, Doppler effects are included. The
model is fairly effective in reproducing spectropolarimetric data of
Seyfert galaxies (see, for example, Young et al. \cite{young99} for
Mrk~509). In particular, it reproduces variations of the
polarization across broad emission lines (Smith et al.
\cite{smith05}). The model is semi-analytical and therefore does not
take multiple scattering into account.

Wolf \& Henning (\cite{wolf99}) present  a Monte-Carlo code used to
compute the polarization obtained by scattering inside axisymmetric
regions. They consider dust and electron scattering for polar
double-cones and equatorial tori. In the Monte-Carlo approach two or
more of such components can be combined and the resulting
polarization spectra are modeled for various inclinations of the
system. Aside from spectropolarimetric modeling, the code by Wolf \&
Henning (\cite{wolf99}) can also produce polarization images, which
are provided for various torus geometries. An important element in
this code is that multiple scattering, which becomes important for
optical depths $> 0.1$, is considered accurately by including the
dependence of the scattering angle and the polarization of a
scattered photon on its incident Stokes vector. For dust scattering,
two different grain size distributions were examined: one
parameterization representing Galactic dust, and the other favoring
larger grains.

Kartje (\cite{kartje95}) also developed a Monte Carlo Code and
investigated quasar schemes with either a torus geometry or conical
stratified winds along the polar axis. In addition to polarization
by scattering, he also considers polarization by dichroic extinction
due to magnetically-aligned dust grains. For a simple unified torus
model he finds that the dominant parameter of the polarization, $P$,
is the torus half-opening angle: for type-2 objects one can find
significant polarization (up to $30\%$) with a position angle
directed perpendicular to the axis of symmetry; for type-1 objects
$P$ is negligible. Kartje obtains an important result when he
investigates conical stratified-wind regions containing free
electrons closer to the central source and dust farther out: the
amount of polarization ranges between $0\%$ and $13\%$, matching
observed values, and the direction of the $\vec{E}$-vector depends
on the viewing angle in a manner that agrees with the
above-mentioned type-1/type-2 dichotomy. The polarization percentage
can be increased if there is magnetic alignment of dust grains, but
the general dependence of $P$ on the viewing angle seems to be a
geometrical effect.

Another Monte-Carlo polarization code is presented by Watanabe et
al. (\cite{watanabe03}). It is applied to modeling of optical and
near-infrared spectropolarimetric data of the type-2 Seyfert galaxies
Mrk~463E, Mrk~1210, NGC~1068, and NGC~4388. The code contains electron
and dust scattering routines quite similar to those used by Wolf \&
Henning (\cite{wolf99}). It considers multiple scattering and dichroic
absorption in dusty tori, spheres as well as electron and dust
scattering in double-conical regions. The absorption and scattering
properties of the dust are carefully calculated by Mie
theory. Watanabe et al. (\cite{watanabe03}) examine
wavelength-dependent polarization properties for different geometries
over a broad-wavelength range and give constrains about possible
scattering components within the objects they observed. They conclude
that a combination of dust and electron scattering in polar regions
can reproduce the optical polarization properties of Mrk~463E
and Mrk~1210. The slope of optical polarization NGC 1068 is almost flat
favoring electron scattering as the dominant polarizing process. For
the near-infrared range polarization of these objects can be modeled
by dichroic absorption of aligned dust grains in a torus. However,
scattering off Galactic dust in a torus cannot simultaneously
reproduce the near-infrared polarization and the total flux. Watanabe et
al.  (\cite{watanabe03}) hence suggest that the grain size
composition of AGNs might be different from our Galaxy.

This list of previous polarization modeling is not exhaustive. For
example, Blaes \& Agol \cite{blaes96} and Agol \& Blaes
\cite{agol96} have presented modeling of the wavelength-dependent
polarization signature of accretion disks at the Lyman limit, and
Kishimoto (\cite{kishimoto96}) modeled polarization due to
electron-scattering off clumpy media in polar regions of AGN. Also, a
new Monte-Carlo model, which includes polarization transfer for the
continuum and for broad quasar absorption lines, was recently
presented by Wang, Wang, \& Wang (\cite{wang06}). We have restricted
this brief review to recent modeling of polarization by dust and
electron scattering, since this is our primary concern in the present
paper.

\section{Stokes -- an overview}
\label{sec:code}

\begin{figure}
 \vskip 0.5cm \resizebox{8cm}{!}{\includegraphics{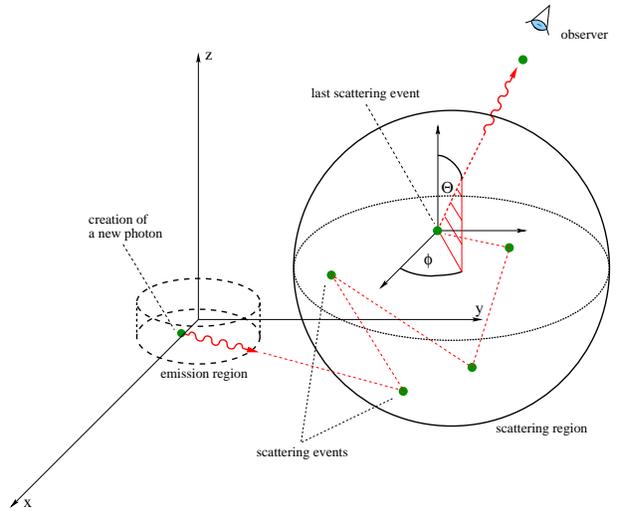}}
 \caption{A photon working its way through the model space.}
 \label{fig:model-space}
\end{figure}

The computer program {\sc Stokes} performs simulations of radiative
transfer, including the treatment of polarization, for AGNs and
related objects. The code is based on the Monte Carlo method and
follows single photons from their creation inside the source region
through various scattering processes until they become absorbed or
manage to escape from the model region (Fig.~\ref{fig:model-space}).
The polarization properties of the model photons are given by their
stored Stokes vectors.

Photons are created inside the source regions, which can be realized
by different geometries. The continuum radiation is normally simply
defined by the index $\alpha$ of an $F_\nu \propto \nu^{-\alpha}$
power law. The Stokes vectors of the photons are initially set to
the values of completely unpolarized light.

Various scattering regions can be arranged around the sources. The
program offers e.g. toroidal, cylindrical, spherical or conical shapes.
These regions can be filled with free electrons or dust consisting
of ``astronomical silicate'' and graphite. A photon works its way
through the model region and generally undergoes several
scatterings. The emission directions, path lengths between
scattering events, and the scattering angles are computed by Monte
Carlo routines based on classical intensity distributions.
During each scattering event the Stokes vector is changed by multiplication
with the corresponding Mueller matrix. For dust scattering, absorption is
important, and a large fraction of the photons never reaches the
virtual observer. The relevant cross sections and matrix elements
for dust scattering and absorption are computed on the basis of Mie
theory applied to size distributions of spherical graphite and
silicate grains.

If a photon escapes from the model region, it is registered by a web
of virtual detectors arranged in a spherical geometry around the
source. The flux and polarization information of each detector is
obtained by adding up the Stokes parameters of all detected photons.
If the model is completely axially symmetric these can be
azimuthally integrated and, if there is plane symmetry, the top and
bottom halves are combined. The object can be analyzed in total
flux, in polarized flux, percentage of polarization, and the
position angle at each viewing angle.  The light travel time of each
photon is also recorded, so it is possible to model time-dependent
polarization (Gaskell, Shoji, \& Goosmann, in preparation).

\subsection{Monte Carlo method, photon initialization, and sampling
the free path length}
\label{sec:routines}

Using the Monte Carlo method it is possible to generate a random
event $x$ according to a given probability density distribution
$p(x)$. Let $p(x)$ be defined on the interval $[0, x_{\rm max}]$. We can
then construct the probability distribution function $P(x)$ and
relate it to a random number, $r$, between $0$ and $1$ as follows:

\begin{equation}
 r = P(x) = \frac{1}{C} \int_0^x p(x')dx'.
 \label{InvInt}
\end{equation}

The constant $C$ is a normalization constant resulting from
integration over the whole definition interval $[0, x_{\rm max}]$. Given
the random number, the corresponding value of $x$ for a single
event is obtained by inverting equation (\ref{InvInt}). A good
description of the Monte Carlo method can be found in Cashwell \&
Everett (\cite{cashwell59}). In the following, we describe the main
routines of {\sc Stokes} and denote all random numbers computed from
equation (\ref{InvInt}) by $r_{\rm i}$, with $i = 1,2,3...$.

To generate a model photon, its initial parameters of position,
direction of flight, and wavelength all have to be set. Different
geometries for the continuum region, broad-line region, and narrow-line
region are available in {\sc Stokes}. Assuming a constant density of the
emitting material, a random position for the new photon is sampled. The
flight direction is given by two angles, $\theta$ and $\phi$, defined
with respect to a standard polar coordinate system. Assuming isotropic
emission, the sampling equations for the angles are as follows:

\begin{eqnarray}
 \theta & = & \arccos(1-2r_1),\\
 \phi & = & 2\pi r_2.
\end{eqnarray}

The wavelength of the photon is sampled according to the intensity
spectrum over a range $[\lambda_{\rm min}, \lambda_{\rm max}]$. This leads to:

\begin{equation}
  \lambda =  \left \{
  \begin{array}{ll}
 \left[ \lambda_{\rm min}^{\alpha} + r_3 \left( \lambda_{\rm max}^{\alpha}
 - \lambda_{\rm min}^{\alpha} \right) \right]^{\frac{1}{\alpha}}, &
 {\rm for} \; \alpha \neq 1,\\\\
 \lambda_{\rm min}
 \left(\frac{\lambda_{\rm max}}{\lambda_{\rm min}}
 \right)^{r_3}, & {\rm for} \; \alpha = 1.
  \end{array}
  \right.
\end{equation}

Here, $\alpha$ denotes the usual power law index of the intensity
spectrum.

If we ignore scatterings back into the beam, the intensity of a
photon beam traversing a slab of scattering material with particle
number density $N$ and cross-section $\sigma$ will drop by a factor
of $e^{N \sigma l}$, with $l$ being the distance traveled inside the
scattering region. From this, one can derive the sampling function of
$l$:

\begin{equation}
 l = \frac{1}{N \sigma} \ln(1-r_4).
\end{equation}

The factor $\frac{1}{N \sigma}$ is the mean free path length.
Depending on the scattering material, the program uses either a dust
extinction cross-section $\sigma_{\rm ext}$ computed from Mie theory or,
in case of electron scattering, the Thomson cross section $\sigma_{\rm ES}$.

\subsection{Polarization formalism, scattering, and photon detection}
\label{sec:P-formalism}

The polarization properties of the photons sampled in {\sc
Stokes} and their transformation during scattering events rely on
previous work described e.g. in Fischer, Henning, \& Yorke
(\cite{fischer94}). The theoretical basis for the formalism
presented in this section can be found in Bohren \& Huffman
(\cite{bohren83}).

\begin{figure}
  \vskip 0.5cm \resizebox{8cm}{!}{\includegraphics{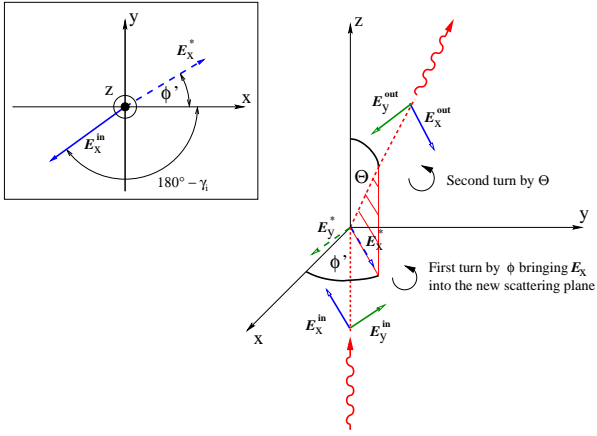}}
  \caption{\label{fig:scattgeom}Geometry and denotations for a single
  scattering event. The inset shows the first rotation of the $\vec
  E$-vector by the angle $\phi$, the view on the polarization plane is
  along the negative $z$-axis.}
\end{figure}

If we consider a photon being scattered off a spherical particle
(see Fig.~\ref{fig:scattgeom}), the outgoing electromagnetic wave
associated with the photon can be resolved into two components, $E_{\parallel}$
and $E_{\bot}$. These components refer to directions of the
electric-field vector parallel and perpendicular to the scattering
plane. For scattering off a spherical particle, the following relation
between the incoming and scattered electric fields holds:

\begin{equation}
 \left(
 \begin{array}{c}
E_{\parallel,{\rm s}}\\
E_{\perp,{\rm s}}
 \end{array}
 \right) =
 \left(
 \begin{array}{cc}
S_2(\theta) & 0\\
0 & S_1(\theta)
 \end{array}
 \right)
 \left(
 \begin{array}{c}
E_{\parallel,{\rm i}}\\
E_{\perp,{\rm i}}
 \end{array}
 \right) .
 \label{eqn:scattmat}
\end{equation}

The scattering matrix elements, $S_1(\theta)$ and $S_2(\theta)$,
are independent of the azimuthal angle $\phi$. In case of Thomson
scattering, there absolute values obey to simple analytic expressions:

\begin{eqnarray}
  |S_1(\theta)|^2 & = & 1,\\
  |S_2(\theta)|^2 & = & \cos^2 \theta.
\end{eqnarray}

For dust scattering, the albedo and the matrix elements of a
standard dust grain are calculated from Mie theory (see section
\ref{sec:mie}). The albedo at the photon wavelength is compared to a
random number, $r_5$, in order to decide whether the photon is
absorbed or scattered. If the photon is absorbed it is lost, and the
cycle starts over with the generation of a new photon.

The polarization vector of each photon lies perpendicular to its
trajectory, inside the so-called polarization plane, and denotes
the preferred direction of the $\vec E$-vector. It is defined with
respect to a co-moving coordinate system. The polarization
information of each photon is coded by four Stokes parameters $I$,
$Q$, $U$, and $V$, representing the 4-dimensional Stokes vector. We
assume that newly created photons coming from the source are
unpolarized. Hence, their Stokes vectors have the simple form:

\begin{equation}
 \left ( \begin{array}{c}  I\\ Q\\ U\\ V
 \end{array} \right )=
 \left ( \begin{array}{c}  1\\ 0\\ 0\\ 0
 \end{array} \right ).
\end{equation}

With each scattering event, the co-moving coordinate system
undergoes a double rotation: the first rotation, by the azimuthal
angle $\phi$, occurs around the current flight direction of the
photon. It rotates the {\vec E}-vector inside the polarization plane
to the position of the new scattering plane (see
Fig.~\ref{fig:scattgeom}). Physically, it does not affect the
polarization state, although the Stokes vector undergoes the following
coordinate transformation.

\begin{equation}
  \left(
 \begin{array}{c}
  I^{*}\\ Q^{*}\\ U^{*}\\ V^{*}
 \end{array}
  \right) =
  \left(
 \begin{array}{cccc}
   1 & 0 & 0 & 0\\
   0 & \cos 2\phi & \sin 2\phi & 0\\
   0 & -\sin 2\phi & \cos 2\phi & 0\\
   0 & 0 & 0 & 1
 \end{array}
  \right)
  \left(
 \begin{array}{c}
  I^{\rm in}\\ Q^{\rm in}\\ U^{\rm in}\\ V^{\rm in}
 \end{array}
  \right).
\end{equation}

The second rotation occurs in the scattering plane by the scattering angle
$\theta$. The change of the Stokes vector is determined by the Mueller
matrix, which for scattering off a spherical particle has the form:

\begin{equation}
  \left(
 \begin{array}{c}
  I^{\rm out}\\ Q^{\rm out}\\ U^{\rm out}\\ V^{\rm out}
 \end{array}
  \right) =
  \frac{1}{k^2d^2}
  \left(
 \begin{array}{cccc}
  S_{11} & S_{12} & 0 & 0\\
  S_{12} & S_{22} & 0 & 0\\
  0 & 0 & S_{33} & S_{34}\\
  0 & 0 & -S_{34} & S_{44}
 \end{array}
 \right)
 \left(
 \begin{array}{c}
   I^{*}\\ Q^{*}\\ U^{*}\\ V^{*}
 \end{array}
 \right).
\end{equation}

The entries of the Mueller matrix are obtained by simple relations
from the elements of the scattering matrix, $S_1(\theta)$ and $S_2(\theta)$.

The angle-dependent classical intensity distribution of a scattered
electromagnetic wave measures the probability of finding a scattered
photon at a given direction. Such probability density distributions
are derived from equation (\ref{eqn:scattmat}). An important aspect
included in {\sc Stokes} is that the scattering direction is sampled
depending on the incident polarization vector. The degree $P_{\rm i}$
and the position angle $\gamma_{\rm i}$ of the polarization before
scattering are computed from the incident Stokes vector and enter the
sampling equations of $\theta$ and $\phi = \phi' + 180^\circ -
\gamma_{\rm i}$:

\begin{eqnarray}
 P(\theta) & = & N \int_{0}^{\theta} \left ( |S_1(\theta')|^2 +
 |S_2(\theta')|^2 \right ) \sin(\theta')d\theta',
 \label{ThetaSamp}\\
 P_{\theta}(\phi') & = & \frac{1}{2\pi}
 \left(\phi'-\frac{|S_1(\theta)|^2-|S_2(\theta)|^2}{|S_1(\theta)|^2+|S_2
 (\theta)|^2} P_{\rm i} \frac{\sin2\phi'}{2} \right).
\end{eqnarray}

The number $N$ is a normalization constant in order to have
(\ref{ThetaSamp}) range from $0$ to $1$ for scattering angles
between $0^\circ$ and $180^\circ$. To sample $\phi$ and $\theta$,
the right hand-sides of these equations have to be set equal to
random numbers $r_6$, $r_7$. The equations are then solved for the
angles.

Note that the sampling is independent of the incident
polarization for $\theta$ but not for $\phi$.
In several Monte-Carlo polarization transfer codes described
in the literature, the incident polarization does not affect the
sampling of the scattering angles. This does not present a problem if
one considers unpolarized incident radiation and low optical depths.
Also for very high optical depths, when multiple-scattering neutralizes
the polarization inside the scattering region, the incident polarization
can be neglected. However, results for intermediate optical depths are
sensitive to the sampling method and they should consider the polarization
state of the incident photon.

When a photon escapes from the model region it is recorded by one of
the virtual detectors. It is then necessary to rotate the polarization
plane around the flight direction until it matches the reference axis
of the detector. The Stokes vectors of all incoming photons can finally
be added up to the values $\hat I$, $\hat Q$, $\hat U$ and $\hat V$. The net
polarization properties are derived from:

\begin{eqnarray}
  P & = & \frac{\sqrt{{\hat Q}^2 + {\hat U}^2 + {\hat V}^2}} {\hat I},\\
  \gamma & = & \frac{1}{2} \arctan\frac{\hat U}{\hat Q}.
\end{eqnarray}

\subsection{Computation of dust properties}
\label{sec:mie}

Mathis, Rumpl, \& Nordsieck (\cite{mathis77}, MRN) suggested dust
compositions to reproduce extinction curves observed in our Galaxy.
They assumed various types of dust grains having a size distribution
proportional to $a^s$, with $a$ being the grain radius and $s$ an
arbitrary power-law index. Our parameterization of the dust
properties follows that of MRN and gives a good description of
observed Galactic extinction curves. The user can choose the
arbitrary minimum and maximum radii of the grain-size distribution,
its power law index, and the relative abundances of graphite and
``astronomical silicate''.

The results from Mie scattering theory, i.e., scattering and
extinction cross sections, albedos, and elements of the scattering
matrix, are computed using the code given by Bohren \& Huffman
(\cite{bohren83}). We imported complex dielectric functions for
graphite and silicate measured by Draine \& Lee (\cite{draine84}).
For graphite, two dielectric functions have to be considered since
the optical properties for light polarized parallel and
perpendicular to the crystals axis differ from each other. The code
therefore works with the two different graphite types having
abundances in a ratio of 1:2. It computes a weighted average
for the dust composition and grain size distribution defined. The
procedure is described, for example, in Wolf (\cite{wolf03}). The
properties of the resulting ``standard dust grain'' are then used by
all dust-related routines of {\sc Stokes}.

\begin{figure}
  \vskip 0.5cm
  \resizebox{8cm}{!}{\includegraphics{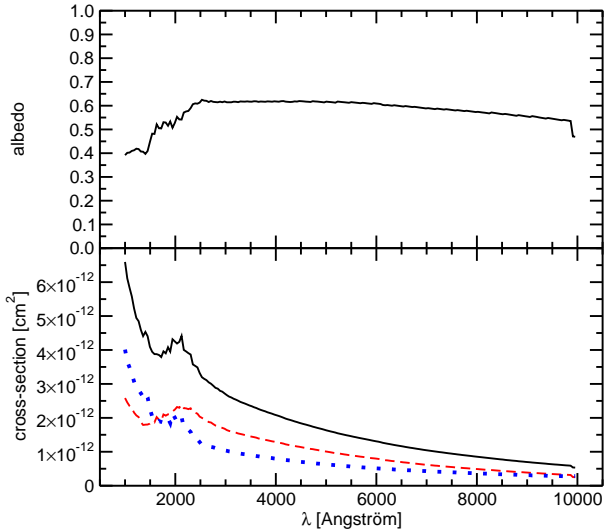}}
  \caption{Characteristic properties of the dust composition adopted
  for this paper as a function of wavelength. Top: albedo
  value. Bottom: cross-sections for extinction (black, solid),
  scattering (red, dashed), and absorption (blue, dotted).}
  \label{fig3}
\end{figure}

We confine ourselves to using standard Galactic dust such as is
seen in the solar neighborhood, even though there is evidence that
the tori of AGNs might have different compositions and grain size
distributions (see Czerny et al. \cite{czerny2004}, Gaskell et al.
\cite{gaskell04}, and Gaskell \& Benker \cite{gaskell06}). Following Wolf \&
Henning \cite{wolf99}, we parameterize Galactic dust by a mixture of 62.5\%
carbonaceous dust grains and 37.5\% ``astronomical silicate''. We consider
grain radii, $a$, from 0.005 $\mu$m to 0.250 ${\rm \mu m}$ with a distribution
$n(a) \propto a^s$ with $s = -3.5$. The resulting cross-sections
and the albedo are shown in Fig.~\ref{fig3} as a function of
wavelength. The figure shows that for this particular dust model, the albedo
is rather flat with a value of 0.55--0.6 over the wavelength range
considered. For wavelengths  $\lesssim 2500$~\AA~it falls to
0.4. The cross-sections all decrease regularly with wavelength,
with the exception of the well-known hump around 2175~\AA.

\section{Simulation of torus geometries}
\label{sec:torus}

In this section we investigate how much of the polarization
properties of type-1 and type-2 AGNs can be produced by a
uniform-density torus alone. Kartje (\cite{kartje95}) modeled the
polarization induced by scattering off a cylindrically shaped torus.
Their torus model was adopted from a fit to NGC~1068 given by Pier
\& Krolik (\cite{pier92}). This torus is geometrically rather
compact and is located within a radius of 1 pc from the central
source. Such a cylindrical torus is not necessarily physical, so we
examine whether the results of Kartje can be confirmed with more
general tori, and we extend the range of parameter space explored.

\begin{figure}
  \vskip 0.5cm
  \centering
  \resizebox{5.5cm}{!}{\includegraphics{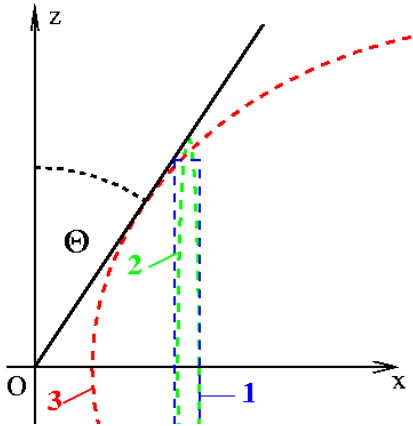}}
  \caption{Geometry of the three torus models we consider: (1) the
 cylindrical torus used by Kartje, (2) a compact elliptically-shaped
 torus, and (3) an extended elliptical torus. All tori have the same
 half-opening angle $\Theta_0$}
  \label{fig4}
\end{figure}

\subsection{Curved surfaces versus sharp edges}
\label{secCurveEdge}

The dusty tori examined by Kartje (\cite{kartje95}), Wolf \& Henning
(\cite{wolf99}), Young (\cite{young00}), and Watanabe et al.
(\cite{watanabe03}) have rather sharp edges, and, since we find that
polarization results can depend strongly on geometrical details, we
have investigated a less artificial torus geometry with an
elliptical cross-section. To examine the influence of sharp edges of
the cylindrical torus on the polarization, we define a torus with
similar dimensions, and the same optical depth in the V band
($\tau_{\rm V}$) $\sim 750$ along the radius in the equatorial plane.
Thus, practically no photon is able to penetrate through the torus
and only scattering off its surface is relevant. However, our torus
has an elliptical cross-section (see Fig.~\ref{fig4}) instead of the
rectangular cross-section used by Kartje.

\begin{figure}
  \vskip 0.5cm
  \resizebox{8cm}{!}{\includegraphics{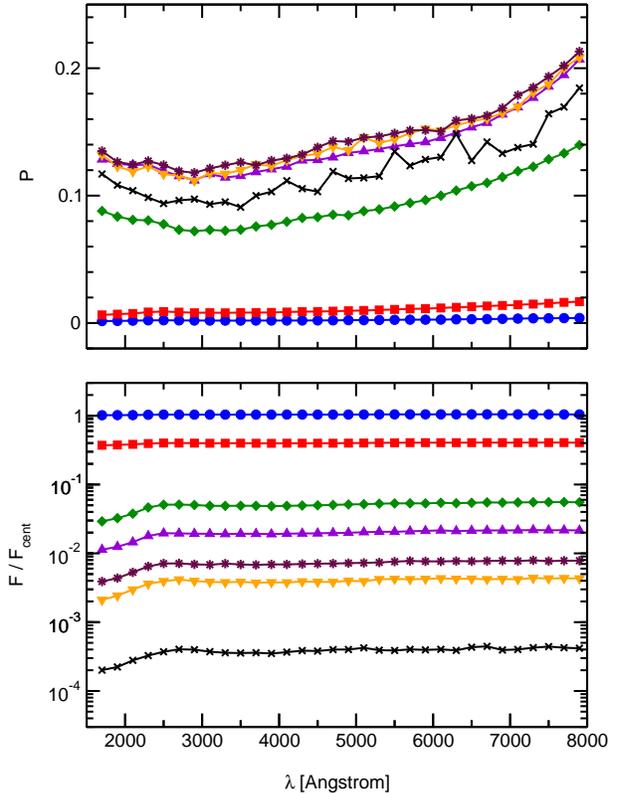}}
  \caption{Modeling a cylindrical torus with an elliptical cross-section
  and $\theta_0 = 30\degr$ (see section \ref{secCurveEdge}). Top:
  polarization, $P$. Bottom: the fraction, $F/F_*$, of the central
  flux, $F_*$, seen at different  viewing inclinations, $i$. Legend:
  $i = 87\degr$ (edge-on)~(black crosses), $i = 76\degr$ (orange
  triangles with points down), $i = 70\degr$ (intermediate)~(maroon
  stars), $i = 57\degr$ (purple triangles with points up), $i =
  41\degr$ (green diamonds), $i = 32\degr$ (red squares), and $i =
  18\degr$ (face-on)~(blue circles).}
  \label{fig5}
\end{figure}

Our results compare very well to those obtained by Kartje
(\cite{kartje95}). In Fig.~\ref{fig5} we show polarization and flux
(normalized to the flux of the central source) versus wavelength at
different viewing directions. The torus considered has a
half-opening angle of $\theta_0=30\degr$. The positive values
of $P$ denote that the polarization vector is oriented
perpendicularly to the symmetry axis (type-2 polarization). In our
simulations the torus is filled with standard Galactic dust,
parameterized as described at the end of section~\ref{sec:code}. We
sample a total of $10^8$ photons and record spectra at $10$
different viewing angles scaled in $\cos i$, where $i$ is measured
from the axis of the torus. We show our results as a function of
$\cos i$ because it gives equal flux per bin for an isotropic source
located at the center of the model space if there is no scattering.
Our figure is quite similar to the corresponding diagrams in
Kartje`s paper (see his Fig.~5).

The only difference between our results and those of Kartje is
that we generally obtain slightly lower polarization degrees and a
slightly different wavelength-dependent slope for the scattered
flux. This can be explained by the fact that we calculate our cross-sections
from Mie theory of a specific dust composition whilst Kartje used
cross-sections given by Mezger, Mathis, \& Panagia (\cite{mezger82}).

We also investigated the polarization of a compact torus with
an elliptical cross-section for changing $\theta_0$. Again we obtained
similar results (not shown) to those for Kartje`s cylindrically-shaped
tori. Thus, the differences in polarization between the elliptical and
cylindrical tori are negligible. Having sharp edges in the cylindrical
model rather than the more realistic rounded edges of the elliptical
torus does not introduce spurious effects.

\subsection{The effect of the shape of the inner edge of the torus}
\label{secBigTorus}

A real torus is undoubtedly thicker than the geometrically thin
cylindrical torus of Kartje. Direct imaging of NGC~4261 (= 3C~270)
shows that the dusty torus in that AGN extends out to 230~pc
(Ferrarese, Ford, \& Jaffe \cite{ferrarese96}). A similar dust lane
across the nucleus of M~51 (= NGC~5194) extends by $\sim$ 100~pc
(Ford et al. \cite{ford92}). The inner radii of tori are obtained by
infra-red reverberation mapping of the hot dust and are in the range
of tens to hundreds of light-days for Seyfert galaxies (see Glass
\cite{glass04} and Suganuma et al. \cite{suganuma06}).

The outer regions of tori have considerable optical depth, so their
precise shape is unimportant, since no photons escape parallel to
the equatorial plane of the torus. The shape of the inner region
facing the central energy source is more relevant. Current torus
models commonly consider inner surfaces that are convex towards the
central source. We thus model optically-thick, uniform-density tori
with elliptical cross sections, an inner radius of $0.25$ pc, and an
outer radius of $100$ pc. We compare these results to the modeling
of a more compact torus with the same half-opening angle,  $\theta_0
= 30\degr$, as in section~\ref{secCurveEdge}. We determine the dust
density by fixing $\tau_{\rm V}$ at $\sim 750$. Variability observations
imply that the size of the optical and UV-continuum source in
Seyfert galaxies is less than a few light-days, as is also expected
from simple black-body emissivity arguments. Hence, when considering
scattering off the torus, we can neglect the finite size of the
continuum source in our model and assume a point-like emission
region. Note that this consideration remains valid for objects
with higher luminosities because both the size of the central
emission region and the inner radius of the torus scale with
luminosity.

\begin{figure}
  \vskip 0.5cm
  \resizebox{8cm}{!}{\includegraphics{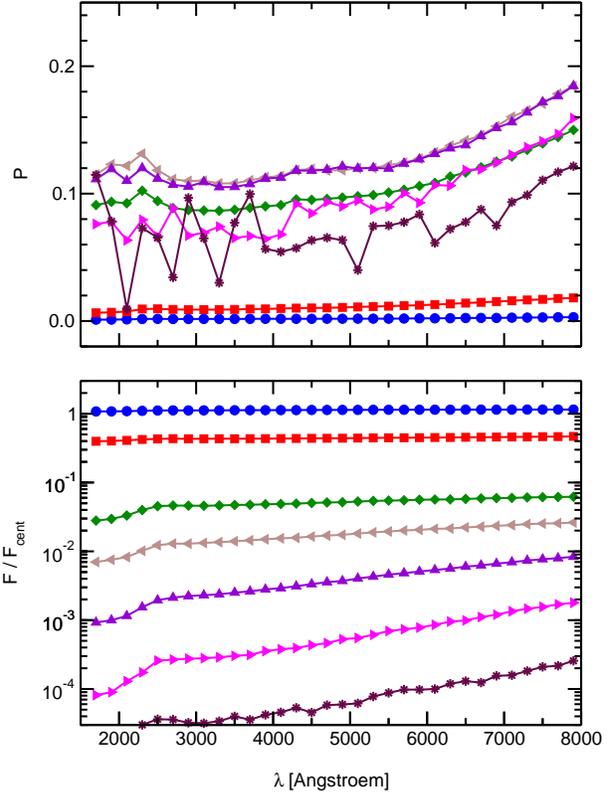}}
  \caption{Modeling a large torus with an elliptical cross-section
 and $\theta_0 = 30\degr$ (see  section \ref{secBigTorus}). Top:
 polarization, $P$. Bottom: the fraction, $F/F_*$, of the central
 flux, $F_*$, seen at different viewing inclinations, $i$. Legend:
 $i = 70\degr$ (intermediate)~(maroon stars), $i = 63\degr$ (pink
 triangles with points to the right), $i = 57\degr$ (purple
 triangles with points up), $i = 49\degr$ (brown triangles with
 points to the left), $i = 41\degr$ (green diamonds), $i = 32\degr$
 (red squares), and $i = 18\degr$ (face-on)~(blue circles)}
  \label{fig6}
\end{figure}

The resulting spectra at different inclinations are shown in
Fig.~\ref{fig6}. If the viewing angle, $i$, is less than $\theta_0$
(thus corresponding to a type-1 object), we only observe a regular
type-1 spectrum. We find that there is no significant polarization
in this case. If we look at a type-2 object at a higher inclination
angle, only scattered (and hence polarized) light is detected. This
is analogous to the results obtained for the compact torus shown in
Fig.~\ref{fig5}. The overall shape of the polarization spectrum for
both sizes of the torus is rather similar as well. With increasing
viewing angle the level of the polarization spectrum rises, reaches
a maximum, and decreases again towards edge-on lines of sight. The
shape of the $P$-spectrum does not change significantly between
different type-2 inclinations.

\begin{figure}
  \vskip 0.5cm
  \centering
  \resizebox{5.5cm}{!}{\includegraphics{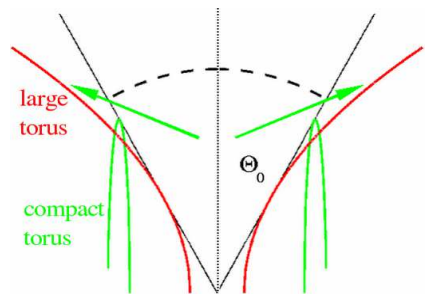}}
  \caption{Comparison of the compact and the extended torus with
  $\theta_0 = 30\degr$ in the V-band.}
  \label{fig7}
\end{figure}

There are differences between our results of modeling a large torus
(case 3 in Fig.~\ref{fig4}) with half-opening angle
$\theta_0=30\degr$, and the analogous compact torus (case 2 in
Fig.~\ref{fig4}) with identical half-opening angle but smaller
dimensions. A striking difference occurs in the angular flux
distribution: the large torus scatters considerably fewer photons
towards an observer at intermediate viewing angles because they hit
the outer parts of it (see the illustration in Fig.~\ref{fig7}).
Towards edge-on viewing directions the probability of seeing
scattered photons is much lower than for the small torus. The
spectral slope of the scattered radiation also differs between the
two tori. While the spectrum is flat in the case of a compact torus
it rises towards the blue for the large torus. This can be explained
by the increasing tendency of forward-scattering at shorter
wavelengths. Photons escaping at higher inclinations have to undergo
back-scattering; this is more likely to happen at longer
wavelengths.

There are also differences in the polarization signatures of both
tori. Although the overall spectral dependence of $P$ is the same,
the level of $P$ is changed. The strongest changes are at higher
inclinations when the central source is becoming obscured by the
torus. As with the total flux (see above), for the larger torus, $P$
is significantly lower (compare the case of $i = 70\degr$ between
the upper panels off Fig.~\ref{fig5} and Fig.~\ref{fig6}). For
a large torus, our current models sampling several $10^9$ photons do
not constrain the polarization well at very high inclinations. The
number of photons scattered into these directions is too small to
allow for sufficient statistics. On the other hand, it clearly
follows from our computations that the spectral flux at angles $i >
76\degr$ is reduced by a factor of almost $\sim 2 \times 10^{7}$
with respect to the flux of the source. Therefore, the polarized flux at these
angles is very low.

\begin{figure}
  \vskip 0.5cm
  \resizebox{8cm}{!}{\includegraphics{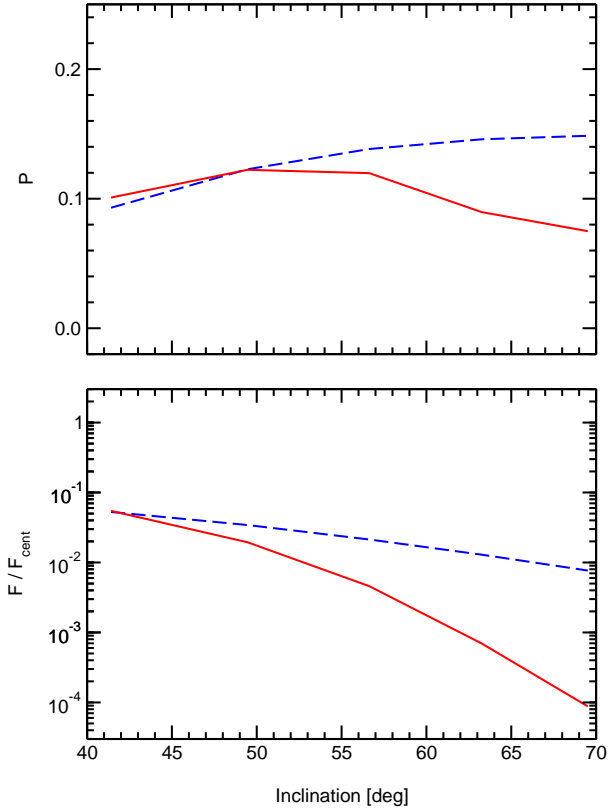}}
  \caption{Differences between large and small tori with an elliptical
 cross-section in the V-band (see section \ref{secBigTorus}). Top:
 polarization, $P$, and bottom, the fraction, $F/F_*$, of the central flux,
 $F_*$, as a function of viewing inclinations, $i$. The dashed lines denote
 the thin elliptical torus (case 2), the solid line the extended torus
 (case 3)}
  \label{fig8}
\end{figure}

We show the differences in V-band total flux and polarization
between a large and a small torus in Fig.~\ref{fig8}. The top panel
shows the polarization as a function of the viewing angle, and the
bottom panel the fraction of the light reaching the observer. As was
shown above, the differences between the two torus shapes are most
important at higher inclinations. At $i \sim 70 \degr$ the degree of
polarization reaches a difference of 6\%, and the flux differs by a
factor of almost $100$.

\subsection{The effect of the torus half-opening angle}
\label{sec:torus-HOA}

Kartje (1995) has shown that the half-opening angle, $\theta_0$, of
the torus is an important parameter for the obscuration and
reflection properties. While modeling large tori, we examine
half-opening angles ranging from $10\degr$ to $75\degr$. Variation
of $\theta_0$ is realized by changing the vertical half-axis of the
elliptical torus cross section. The other model parameters are
defined as for the previous case of $\theta_0 = 30\degr$ in
section~\ref{secBigTorus}.

\subsubsection{Tori with narrow or wide openings}

\begin{figure}
  \vskip 0.5cm
  \resizebox{8cm}{!}{\includegraphics{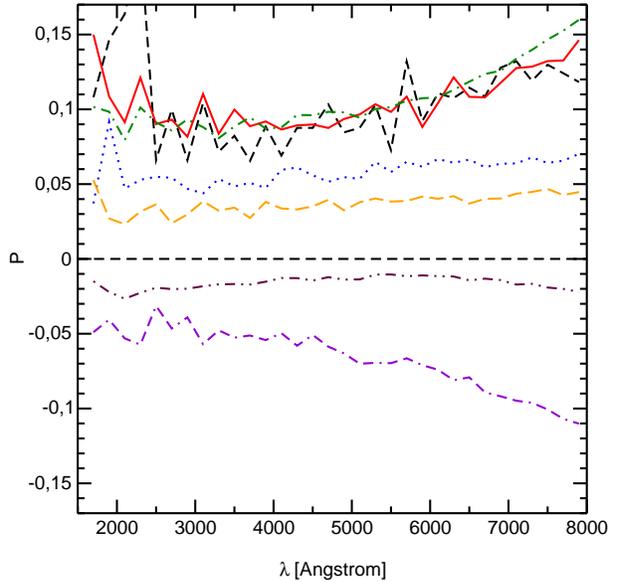}}
  \caption{Polarization averaged over type-2 viewing angles (see
  section~\ref{sec:torus-HOA}). A positive value of polarization
  denotes an $\vec{E}$-vector oriented perpendicular to the torus
  symmetry axis; for negative values the $\vec{E}$-vector is aligned
  with the projected axis.  Legend: $\theta_0 = 10\degr$ (black dashed
  line), $\theta_0 = 20\degr$ (solid red line), $\theta_0 = 30\degr$
  (green dot-dashed line),  $\theta_0 = 45\degr$ (blue dots), $\theta_0 =
  50\degr$ (long yellow dashes), $\theta_0 = 60\degr$ (brown double dots
  and dashes), and $\theta_0 = 75\degr$ (pink double-dashes and
  dots). \label{fig9}}
\end{figure}

For large tori, $\theta_0$ is a dominant parameter for both the
degree of polarization and the position angle, $\gamma$. In
Fig.~\ref{fig9} we show the polarization of the scattered radiation
as a function of wavelength and for various $\theta_0$. Due to
a similar overall shape of the wavelength dependence of $P$, we
average the polarization over type-2 viewing angles, $i$, with $i >
\theta_0$. We thereby exclude the highest inclinations with an
insufficient number of photons, where the statistics of $P$ are too
poor. For viewing angles with $i < \theta_0$ (corresponding to
type-1 objects seen face-on) the polarization is negligible.

Varying the opening angle shows several important things. For
$\theta_0 < 53\degr$ the absolute value of the polarization
decreases as the opening angle increases (see Fig.~\ref{fig9}), as
was found by Kartje (\cite{kartje95}) for compact tori. The polarization
vector is oriented perpendicularly to the axis for all viewing directions $i >
\theta_0$, as is observed in type-2 AGN. For $\theta_0 > 60\degr$, only
parallel polarization vectors can be seen at viewing angles $i > \theta_0$. In
this range of $\theta_0$ the absolute degree of polarization increases with
the opening angle.

The reason for the flip of the relative position angle can be
explained by the scattering phase function, and by the geometry of
the inner parts of the torus (Kartje \cite{kartje95}). For a distant
observer looking at the torus along an off-axis line of sight, the
scattered radiation comes from the inner surface walls. In part,
these consist of the inner torus wall facing the observer most
directly, but they also consist of the two surfaces on the side. Due
to the scattering geometry, the photons scattered off the side walls
are polarized along the projected symmetry axis, whilst the photons
coming from the far wall are perpendicularly polarized. The ratio of
the solid angle that the far side of the visible inner surface
subtends to the solid angle that the visible inner side walls
subtend changes with the half-opening angle of the torus, and so
does the overall polarization vector.

\subsubsection{Transition case: intermediate torus half-opening angles}

\begin{figure}
  \vskip 0.5cm
  \resizebox{8cm}{!}{\includegraphics{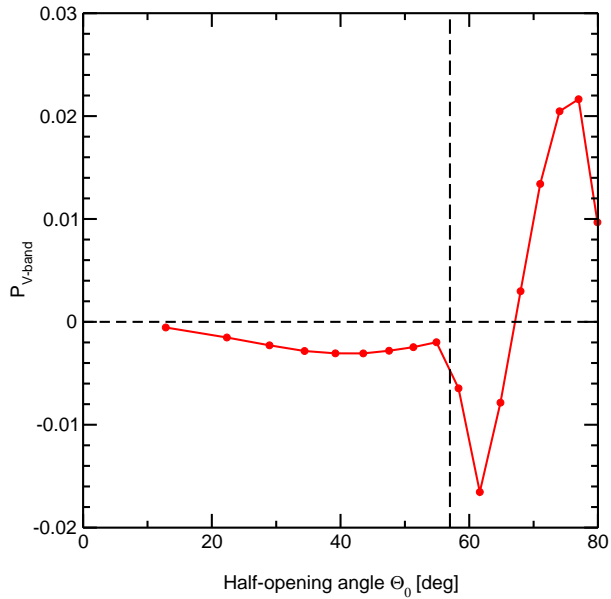}}
  \caption{Modeling an expanded torus at an intermediate opening angle
  of $\theta_0 = 57\degr$. The graph shows the polarization degree in
  the visual band versus inclination angle $i$. A positive value of
  polarization denotes an $\vec{E}$-vector oriented perpendicular to
  the torus symmetry axis; for negative values the $\vec{E}$-vector is
  aligned with the projected axis.}
  \label{fig10}
\end{figure}

For intermediate opening angles with $53\degr < \theta_0 <
60\degr$ the orientation of the polarization position angle seen at
type-2 viewing angles depends on the exact inclination. We
illustrate such a case in Fig.~\ref{fig10}, where we set $\theta_0 =
57\degr$.

For a line of sight passing close enough to the horizon of the
torus (i.e., when $i$ is only moderately larger than $\theta_0$) we
find that the polarization vector is parallel, which means that
type-1 polarization can be produced at obscured viewing inclinations
(Fig.~\ref{fig10}). If the inclination increases further the
polarization vector switches back to type-2 polarization. It is
interesting to note that such a torus can produce significant
polarization degrees up to 2\% for both orientations of the $\vec
E$-vector.

\begin{figure}
 \vskip 0.5cm
 \resizebox{8cm}{!}{\includegraphics{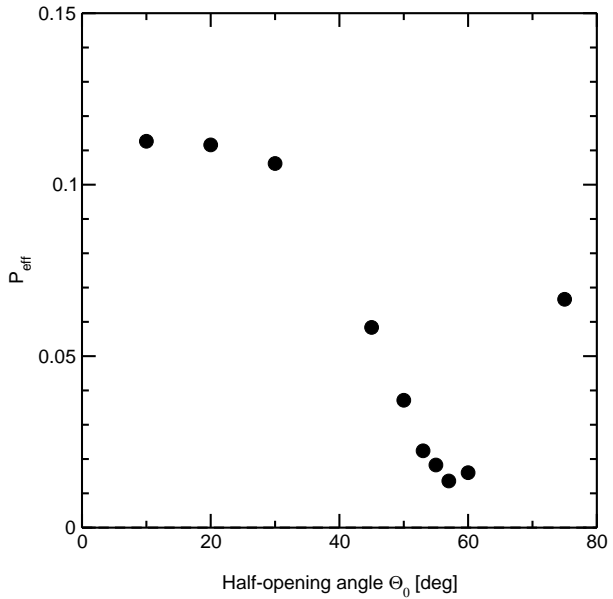}}
 \caption{Effective polarization, $P_{\rm eff}$ (see
 section~\ref{sec:torus-HOA}), for type-2 viewing positions as a
 function of the half-opening angle of an extended torus (case
 3). \label{fig11}}
\end{figure}

In order to illustrate the integral effect of the opening angle on
the polarization, we plot in Fig.~\ref{fig11} the polarization,
$P_{\rm eff}$, averaged over all type-2 viewing positions, and over
wavelength, as a function of the half-opening angle of the torus.
The difference between type-1 and type-2 polarization is ignored in
$P_{\rm eff}$. The absolute values of $P$ are integrated.

The figure shows that the torus polarizes most effectively when
having either a small or a large half-opening angle. In the
transition region between type-1 and type-2 polarization (i.e. for
$53\degr < \theta < 60\degr$) the integrated polarization goes
through a minimum.

\subsection{Wavelength insensitivity of polarization due to dust
scattering}

Wavelength-independent polarization is widely taken to be a
signature of electron scattering, but we have shown in Figs. 6 and 9
that dust scattering can also produce wavelength-independent
scattering. Thus a flat polarization curve is {\it not} a unique
signature of electron scattering. As Zubko \& Laor (\cite{zubko00})
point out, the wavelength dependence of polarization provides a
probe of the grain scattering properties. Inspection of the
wavelength-dependent polarization curves for the large torus
geometries considered above (see Figs. 6 and 9) shows that the
polarization for half-opening angles with $30\degr < \theta_0 <
60\degr$ is wavelength-independent over the optical and most of the
UV. For values of $\theta_0$ outside this interval the
wavelength-dependence of $P$ is rather low and does not exceed a
factor of 2. Since the scattering cross section of interstellar
grains increases strongly from the optical to the UV,
wavelength-independent polarization is commonly supposed to be the
fingerprint of electron scattering. However, scattering in
opaque dust clouds produces relatively grey scattering (Kishimoto
\cite{kishimoto01}).

Our apparently contradictory result of relatively wavelength
independent polarization with dust scattering arises because we are
considering scattering off optically-thick material, and because of
the relatively small variation of the albedo over the optical and UV
spectral regions (see Fig.~\ref{fig3}). The approximate constancy of
the albedo is because the scattering and absorption cross-sections
vary in a similar manner with wavelength. Since we assume an
optically-thick torus, we see emergent photons that have been
scattered at an optical depth $\tau \sim 1$. This is regardless of
wavelength\footnote{This is the reason that the sunlit sides of
clouds in the earth`s atmosphere are extremely white.}. The increase
in scattering cross-section with decreasing wavelength only means
that the shorter wavelength photons we see have been scattered
closer to the surface of the torus.

A significant change in albedo with wavelength, however, will cause
a color dependency in the intensity and polarization of the
scattered light\footnote{This is the cause of colorations in the
atmosphere of the giant planets.}. Shortwards of $\sim
2500$~\AA~the albedo decreases, but this range is at the lower limit
of the spectral range considered in our modeling. The effect can be
seen in the normalized flux spectra of the torus models shown in
Figs.~\ref{fig5} and \ref{fig6}. For the Galactic dust composition
we implemented, it is less visible in the polarization spectra.

Another grain property that needs to be considered is the degree of
asymmetry of the scattering since this is effectively an
angle-dependent albedo change. Toward shorter wavelengths,
Galactic dust grains are more strongly forward scattering and the
polarization phase function changes (see Draine \cite{draine03}). As
for Thomson scattering, forward-scattered light has a lower
polarization than sideways-scattered light. The polarization spectra
obtained for the torus models depend on these phase functions. They
additionally explain why a slight wavelength-dependence of the
polarization can be found for very narrow or very wide opening
angles of a large torus (see Fig.~\ref{fig9}).

\section{Polarization from polar-scattering regions}
\label{sec:cones}

Scattering in polar regions of AGNs has allowed the discovery
of hidden Seyfert-1 nuclei in type-2 objects by radiation being
periscopically scattered around the obscuring torus. The central
parts of the polar double cone have to be at least moderately
ionized due to the  intense radiation from the AGN. The medium could
be associated with the warm absorber seen in many AGN (see Komossa
\cite{komossa99} for a review). The Doppler shift of the X-ray
absorption lines indicates that the medium is outflowing at roughly
1000 km/s. With increasing distance from the center, the outflow
velocity and intensity of the radiation decrease. Beyond the
sublimation radius, dust could also be present. However, this dust
must be optically thin, as type-1 objects are not obscured.

\subsection{Polar electron scattering}

The polarization induced by scattering in polar, conical
electron-scattering regions has been the subject of several previous
studies. Brown \& McLean (\cite{brown77}) developed a formalism to
compute the polarization expected from scattering inside optically
thin, axisymmetric scattering regions. This formalism was applied
by Miller \& Goodrich (1990) and Miller et al. (1991) to compute the
polarization for polar scattering cones. Wolf \& Henning (1999) and
Watanabe at al. (2003) extended the modeling to optically-thick
material using Monte-Carlo techniques that can account for multiple
scatterings.

\begin{figure}
  \vskip 0.5cm
  \resizebox{8cm}{!}{\includegraphics{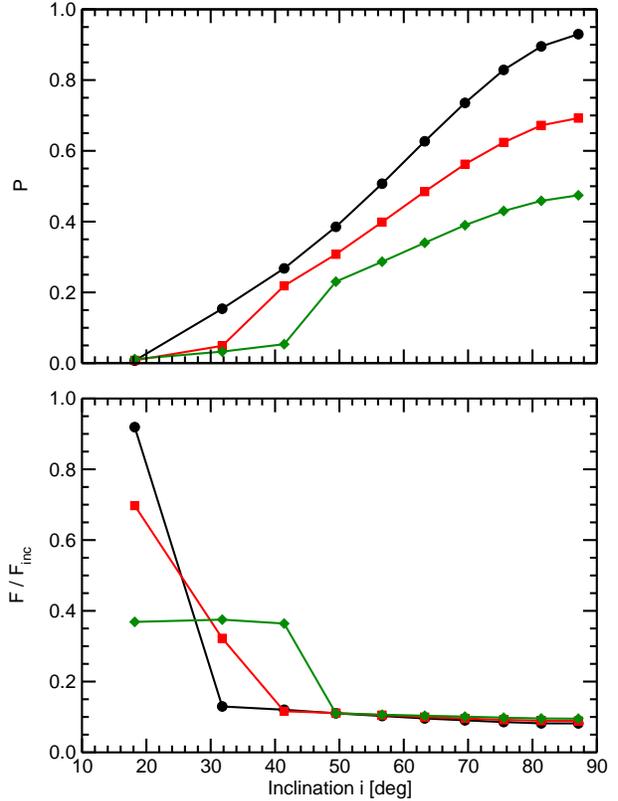}}
  \caption{Modeling double polar cones of various half-opening
  angles. Top: polarization, $P$, with positive values denoting type-2
  polarization (perpendicular to the symmetry axis). Bottom: the
  fraction, $F/F_*$, of the central flux. Both values are plotted
  versus the inclination $i$ with respect to the observer. The
  different symbols denote different half-opening angles of the
  double-cone. Legend: $\theta_{\rm C} = 10\degr$ (black circles),
  $\theta_{\rm C} = 30\degr$ (red squares), and $\theta_{\rm C} =
  45\degr$ (blue diamonds). The optical depth between the inner and
  outer shell of the cones is set to $\tau_{\rm es} = 1$.}
  \label{fig12}
\end{figure}

We confirm such results in Fig.~\ref{fig12} using {\sc Stokes}. The
figure shows the degree of polarization and the total flux as a
function of the observer`s inclination for an electron scattering
double-cone of uniform density and with the optical depth $\tau_{\rm es}
= 1$. This optical depth is measured in the vertical direction
between the inner and the outer shell of one cone.  In order to
isolate the effects of the scattering cone from the polarization
induced by the disk, we use an anisotropically emitting central
source with the emission angles being restricted to the solid angle
defined by the scattering cones.  The three curves denote the
half-opening angles $\theta_{\rm C} = 10\degr$, $\theta_{\rm C} =
30\degr$, and $\theta_{\rm C} = 45\degr$. The inclination is measured
from the symmetry axis of the double-cone.

As expected, polar electron-scattering cones produce type-2
polarization directed perpendicularly to their symmetry axis. The
degree of polarization rises from face-on to edge-on viewing angles.
The latter effect is due to the angle-dependent polarization phase
function of Thomson scattering. For wider opening angles of the
cones, the net polarization $P$ decreases because it is the result
of integrating a broader distribution of polarization vectors. The
breaks of the polarization curves at $i = \theta_{\rm C}$ in
Fig.~\ref{fig12} are due to the impact of  multi-scattering inside
the cones, the analogous breaks in total flux curves are due to the
angle-restricted central emission.

\begin{figure}
  \vskip 0.5cm
  \resizebox{8cm}{!}{\includegraphics{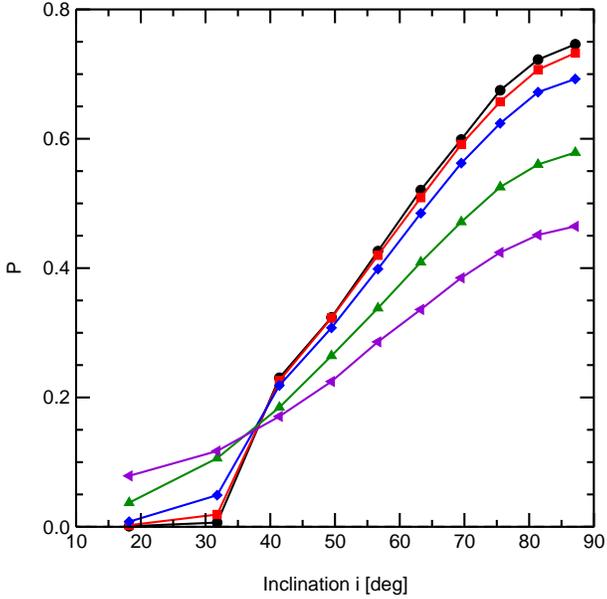}}
  \caption{Polarization degree by polar electron scattering cones with
  half-opening angle $\theta_{\rm C} = 30\degr$ plotted versus the
  inclination $i$ with respect to the observer. The positive values
  denote type-2 polarization (perpendicular to the symmetry axis). The
  different symbols denote different optical depths between the inner
  and the outer radius of the cone. Legend: $\tau_{\rm ES} = 0.01$
  (black circles), $\tau_{\rm ES} = 0.3$ (red squares), $\tau_{\rm ES}
  = 1$ (blue diamonds), $\tau_{\rm ES} = 3$ (green triangles with
  points up), and $\tau_{\rm ES} = 10$ (triangles with points left).}
  \label{fig13}
\end{figure}

In Fig.~\ref{fig13} we plot the influence of the optical depth on the
polarization for the polar-cones with $\theta_{\rm C} = 30\degr$. The
various curves denote different optical depths. A similar case was
considered by Watanabe et al. (\cite{watanabe03}). The density of
their electron cones varies with the distance from the center
according to a power law. Comparison of Fig.~\ref{fig13} with Fig.~7
of Watanabe et al. (\cite{watanabe03}, bottom panel) shows that the
difference in $P$ is very small with respect to a uniform-density
torus.

\subsection{Polar dust scattering}

\begin{figure}
  \vskip 0.5cm
  \resizebox{8cm}{!}{\includegraphics{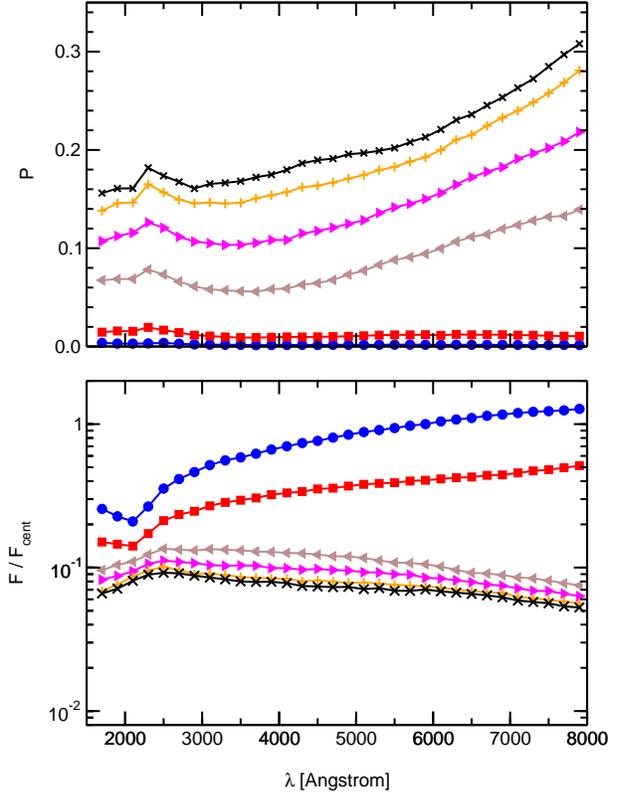}}
  \caption{Modeling a dusty double cone of half-opening angle
 $\theta_{\rm C} = 30\degr$. Top: polarization, $P$. Bottom: the
 fraction, $F/F_*$, of the central flux, $F_*$, seen at different
 viewing inclinations, $i$. Legend: $i = 87\degr$ (edge-on, black
 crosses), $i = 76\degr$ (orange pluses), $i = 63\degr$ (pink
 triangles with points to the right), $i = 49\degr$ (brown
 triangles with points to left left), $i = 32\degr$ (red squares),
 and $i = 18\degr$ (face-on, blue circles).}
  \label{fig14}
\end{figure}

Beyond the dust sublimation radius the scattering cone could contain
dust. We investigated this using a similar bi-conical geometry and our
Galactic dust prescription -- see section~\ref{sec:code}. In
Fig.~\ref{fig14} we show the polarization and total flux resulting
spectra for a centrally-illuminated dust cone seen at different
viewing angles $i$. The half-opening angle of the cone has been set to
$\theta_{\rm C} = 30\degr$, and its optical depth in the V-band to the
moderate value of $\tau_{\rm V} = 0.3$. The strong wavelength
dependence of the dust extinction properties (see Fig.~\ref{fig3}) is
clearly visible in the figure. It differs for polar viewing angles,
which cross the cone, from those along equatorial directions. The
former ones show the dust extinction seen in transmission, while the
latter ones show dust reflection.

The total flux shown in the bottom panel of Fig.~\ref{fig14} is
significantly reddened when $i < \theta_{\rm C}$. In addition to that, the
well known extinction feature at $2175 \mu$m is seen. Its depth
decreases with increasing inclination. In reflection (i.e., along
equatorial viewing angles) the spectra are quite different. They
peak at $\sim 2500$~\AA, where the albedo goes through a maximum
(see Fig.~\ref{fig3}), and level down slowly toward longer
wavelengths. This behavior is the same at all inclinations with $i >
\theta_{\rm C}$; the spectra differ just slightly in normalization.

The polarization spectra remain below 1\% when the double-cone
is seen in transmission because forward scattering does not induce
significant polarization. In reflection, however, polarization
becomes significant, and, for all inclinations with  $i > \theta_{\rm C}$,
it rises towards longer wavelengths. The shapes of the spectra are,
again, not very dependent on the inclination; they only differ in
normalization. The highest level of polarization is obtained when
the cone is seen edge-on, which corresponds to perpendicular
scattering angles.

Note that, while the wavelength dependence in polarization and
flux is different from the one obtained for the dusty torus
investigated in section~\ref{sec:torus}~(because we are considering
optically thin material in the cones), the of the flux rise to shorter
wavelengths is still less than the rise in the cross sections.

\section{Polarization from equatorial electron distributions}

The major difficulty in modeling the type-1/type-2 polarization
dichotomy is producing polarization {\it parallel} to the symmetry
axis. We have illustrated above that polar scattering cones can only
produce polarization vectors oriented {\it perpendicular} to the
axis, and that the polarization produced by the dusty torus at
type-1 viewing angles is mostly negligible. An exception was the
case of extremely-thin tori, but these are inconsistent with the
observations for at least two reasons: they are incompatible with
the ratios of type-1 and type-2 objects, and very flat distributions
have too low covering factors to produce the reprocessed IR
emission.

It is therefore necessary to introduce a third type of scattering
region. It was pointed out a long time ago that parallel polarization
can be produced by a thin emitting and scattering disk (e.g.,
Chandrasekhar \cite{chandrasekhar60}, Angel \cite{angel69}, Antonucci
\cite{antonucci84}, Sunyaev \& Titarchuk \cite{sunyaev85}). Goodrich
\& Miller (\cite{goodrich94}) suggested that the parallel polarization
of type-1 objects arises from scattering in a flattened equatorial
medium located around the accretion disk or even the BLR. This model
has been investigated in a series of papers by Young \cite{young00},
Smith et al.  \cite{smith02},~\cite{smith04}, and \cite{smith05})
using the GSM. In this model a rotating equatorial scattering disk
explains the velocity-dependent polarization structure across the
broad emission lines of a large sample of Seyfert galaxies very well,
and also helps explaining the type-1/type-2 dichotomy. Unlike the GSM,
our code includes the effects of multiple scattering, so we can
consider scattering regions with higher optical depths. We first
constrain possible contributions of an AGN accretion disk to the
polarization (\ref{sec:emiscattdisk}). Then, we investigate with {\sc
Stokes} the effect of equatorial scattering disks  on the continuum
radiation (section~\ref{sec:annuli}).

\subsection{Emitting and scattering disks}
\label{sec:emiscattdisk}

\begin{figure}
  \vskip 0.5cm
  \resizebox{8cm}{!}{\includegraphics{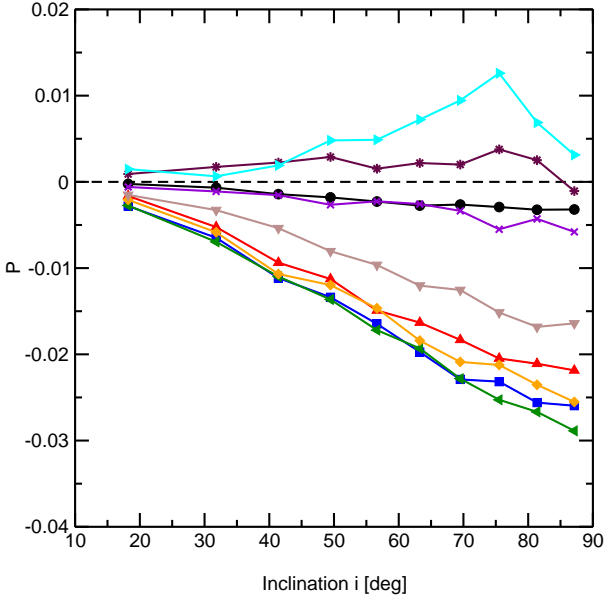}}
  \caption{Polarization versus viewing angle for a geometrically-thick
 emitting and scattering electron disk with $\frac{h}{d}=0.5$. The
 different curves denote various vertical optical depths for the
 disk. Positive polarization values stand for polarization position
 angles perpendicular to the disk`s symmetry axis, for negative values
 the polarization vector is aligned with this direction. Legend:
 $\tau_{\rm ES}$ = 0.05 (black circles), $\tau_{\rm ES}$ = 0.5 (red
 triangles with points up), $\tau_{\rm ES}$ = 1.25 (blue squares),
 $\tau_{\rm ES}$ = 1 (green triangles with points left), $\tau_{\rm
 ES}$ = 1.5 (orange diamonds), $\tau_{\rm ES}$ = 2.5 (brown triangles
 with points down), $\tau_{\rm ES}$ = 5 (purple crosses), $\tau_{\rm
 ES}$ = 10 (maroon stars), and $\tau_{\rm ES}$ = 25 (cyan triangles
 with points right).}
  \label{fig15}
\end{figure}

\begin{figure}
 \vskip 0.5cm
 \resizebox{8cm}{!}{\includegraphics{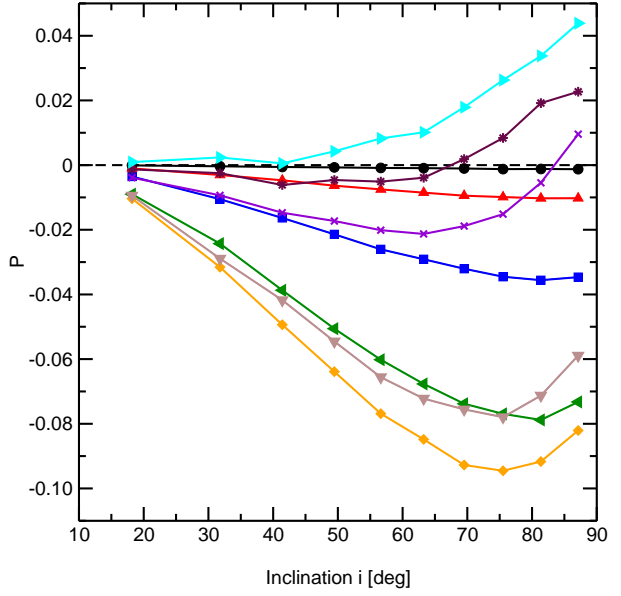}}
 \caption{Same parameters as in Fig.~\ref{fig15}, but for a thin disk with
 $\frac{h}{d}=0.01$. Legend: $\tau_{\rm ES}$ = 0.001 (black circles),
 $\tau_{\rm ES}$ = 0.01 (red triangles with points up), $\tau_{\rm ES}$ = 0.05
 (blue squares), $\tau_{\rm ES}$ = 0.2 (green triangles with points left),
 $\tau_{\rm ES}$ = 0.5 (orange diamonds), $\tau_{\rm ES}$ = 1 (brown triangles
 with points down), $\tau_{\rm ES}$ = 3 (purple crosses),  $\tau_{\rm ES}$
 = 5 (maroon stars), and $\tau_{\rm ES}$ = 10 (cyan triangles with
 points right).}
 \label{fig16}
\end{figure}

We consider emitting and scattering disks with a ratio
$\frac{h}{d}$ of total disk height $h$ to diameter $d$. The disk is
modeled in a plane-parallel approximation. It has a cylindrical
shape and is uniformally filled with electrons. We consider a ``thick
disk'' cross section with $\frac{h}{d}=0.5$ (see Fig.~\ref{fig15})
and a ``thin disk'' with $\frac{h}{d}=0.01$ (see Fig.~\ref{fig16}).
The various curves in the diagrams refer to different vertical
electron-scattering optical depths, $\tau_{\rm ES}$, measured vertically
from the central plane to the surface. The optical depth is varied
by adjusting the electron density. We investigate $\tau_{\rm ES}$
between $0.001$ and $50$.

For the case of a geometrically-thick emitting disk (see Fig.
\ref{fig15}) we find moderate polarization values of at most a few
percent. The degree of polarization depends strongly on the viewing
direction and the optical depth. For lower optical depth the
$\vec{E}$-vector is aligned with the disk`s symmetry axis. It flips
to a perpendicular orientation for $\tau_{\rm ES}$ greater $\sim 10$.

Much stronger values of the polarization can be obtained when the
disk is flatter. For the geometrically-thin disk we have a similar
qualitative behavior as for the thick disk (compare Fig.~\ref{fig16}
with Fig.~\ref{fig15}), but the degree of polarization reaches higher
values and is significant even for near to face-on viewing
directions. The flip to perpendicular (type-2) polarization occurs at
a moderate optical depth ($\tau_{\rm ES} \sim 5$)  and at an edge-on
viewing angle that cannot be seen for type-1 objects. A thin disk with
moderate optical depth will thus produce parallel polarization for all
type-1 viewing positions.

The polarization behavior of our uniformly emitting disks can be
explained in the same manner as the behavior of the polarization of
the oblate spheroids examined by Angel (\cite{angel69}). For low
$\tau_{\rm ES}$ the net polarization is mainly determined by the photons
traveling parallel to the disk plane and then being scattered
towards the surface. For an observer, who does not observe the disk
exactly face-on, the integrated scattered flux from the disk surface
will be polarized along the projected direction of the disk axis.
This can be understood by the fact that polarization by electron
scattering is most efficient for orthogonal scattering angles. For
the same reason, the polarization is also strongest at edge-on
viewing angles.

When the optical depth becomes higher, multiple scattering of
photons traveling toward the disk surface becomes relevant. The
polarization vector induced by the last scattering event before
leaving the disk will preferably be oriented perpendicular to the
disk axis. Hence, the polarization position angle flips and on the
way to this transition $P$ becomes very low.

The emission and scattering disks investigated in this section
are unlikely the cause of type-1 polarization in AGN. High enough
polarization parallel to the disk symmetry axis is only produced in
a disk which is geometrically {\it and} optically thin. While
accretion disks according to the standard model of Shakura \&
Sunyaev (1973) are indeed geometrically thin, they are, however,
optically thick. They produce too little polarization at type-1
viewing angles, as one can tell from the curves in the left part of
Fig.~\ref{fig16}. At type-2 angles the polarization can reach up to the
well-known limit of $\sim 11.7\%$ for the highest optical depths
(Chandrasekhar \cite{chandrasekhar60}), but with a perpendicular orientation
of the polarization vector.

\subsection{Equatorial scattering wedges}
\label{sec:annuli}

Instead of scattering from the accretion disk, it has been
proposed that type-1 polarization is caused by scattering by an
equatorial wedge (Goodrich \& Miller \cite{goodrich94}) or by a
flared equatorial disk (Smith et al. \cite{smith02}). Here, we use
{\sc Stokes} to model such equatorial regions for different
geometrical shapes and optical depths.

\begin{figure}
\vskip 0.5cm \centering
\resizebox{7cm}{!}{\includegraphics{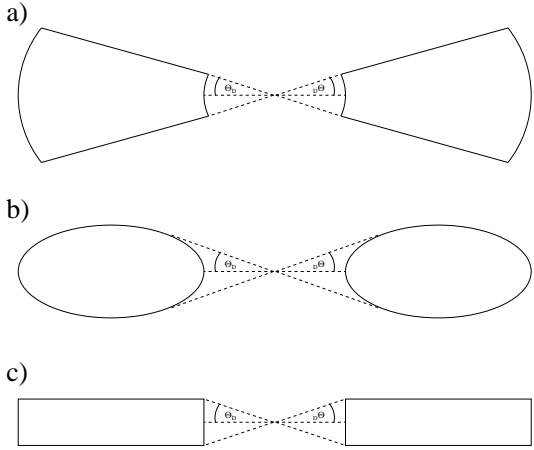}} \caption{The
equatorial scattering geometries investigated: (a) flared disks,
(b) tori, and (c) equatorial cylinders. The definition of the
half-opening angle $\theta_{\rm D}$ is shown for each case.} \label{fig17}
\end{figure}

\subsubsection{Flared disks}
\label{sec:flared-disks}

The geometrical shape of a flared disk of half-opening angle
$\theta_{\rm D}$ is given in Fig.~\ref{fig17} (a). We assume that the
scattering is only off electrons. This is a good assumption for AGNs
if the scattering region is located close enough to the center for
hydrogen to be mostly ionized. We fix the Thomson optical depth in
the equatorial plane between the inner and the outer shell,
$\tau_{\rm ES}$, to have $\tau_{\rm ES} = 1$.

\begin{figure}
  \vskip 0.5cm
  \resizebox{8cm}{!}{\includegraphics{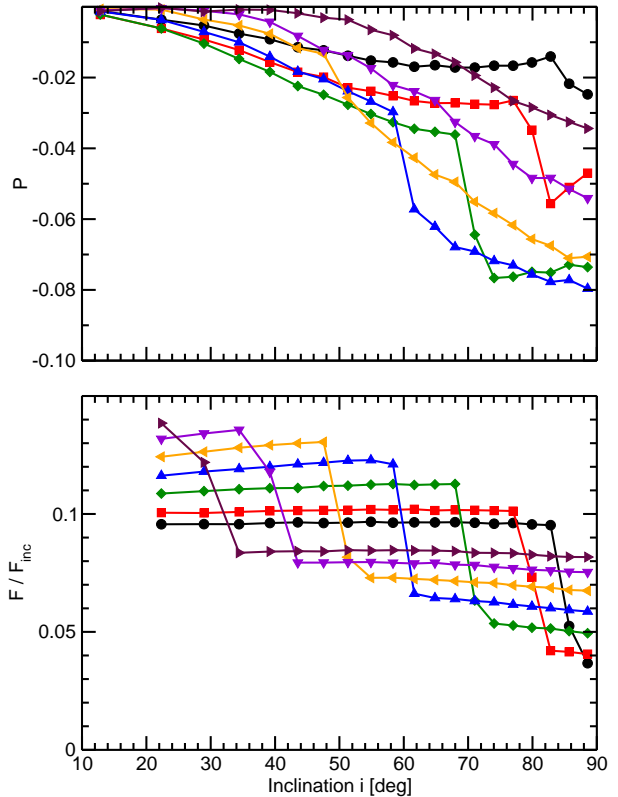}}
  \caption{Modeling equatorial flared scattering disks of various
  half-opening angles $\theta_{\rm D}$ (see section
  \ref{sec:flared-disks}). Top: polarization, $P$, with negative
  values denoting type-1 polarization (parallel to the symmetry
  axis). Bottom: the fraction, $F/F_*$, of the central flux, $F_*$,
  plotted versus inclination, $i$, with respect to the observer. The
  different symbols denote different half-opening angles of the flared
  disk. Legend: $\theta_{\rm D} = 5\degr$ (black circles),
  $\theta_{\rm D} = 10\degr$ (red squares), $\theta_{\rm D} = 20\degr$
  (green diamonds), $\theta_{\rm D} = 30\degr$ (blue triangles with
  points up), $\theta_{\rm D} = 40\degr$ (orange triangles with points
  to the left), $\theta_{\rm D} = 50\degr$ (purple triangles with
  points down), and $\theta_{\rm D} = 60\degr$ (brown triangles with
  points to the right).}
  \label{fig18}
\end{figure}

In Fig.~\ref{fig18}, we plot the total flux and polarization
versus inclination, $i$, for flared disks with various half-opening
angles. Both the total flux and the polarization depend
significantly on the half-opening angle of the flared disk, but the
polarization is always parallel to the symmetry axis of the system.
The flared disk hence produces type-1 polarization, while the
polarization degree obtained at type-1 viewing angles remains rather
moderate at around 2\%.

For viewing angles $i < 90\degr - \theta_{\rm D}$, the central source is
directly visible, which leads to higher fluxes than towards edge-on
viewing angles $i > 90\degr - \theta_{\rm D}$. If the scattering wedge
crosses the line of sight, a fraction of the radiation is scattered
out of the way. The polarization degree increases as the inclination
goes from face-on towards edge-on viewing angles. At face-on
inclinations, the line of sight is nearly aligned with the axis of
the wedge and therefore the observer sees a more axisymmetric
system.

\begin{figure}
  \vskip 0.5cm
  \resizebox{8cm}{!}{\includegraphics{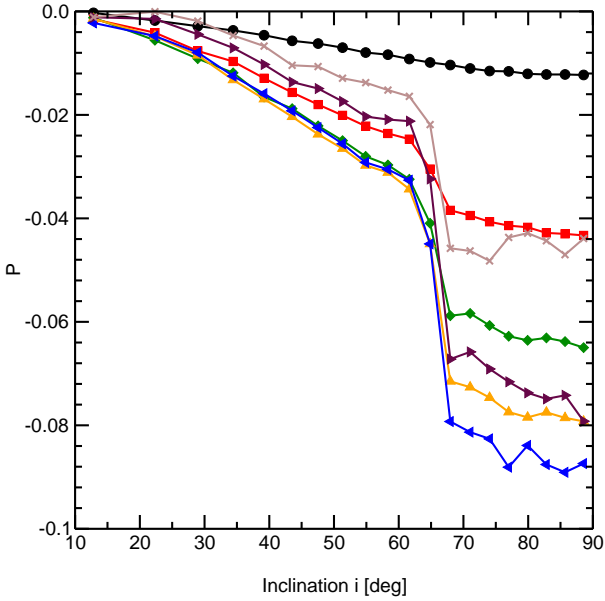}}
  \caption{The degree of polarization by electron scattering by an
  equatorial flared disk with half-opening angle $\theta_{\rm D} =
  25\degr$ plotted versus the inclination, $i$, with respect to the
  observer. The negative values denote type-1 polarization (parallel
  to the symmetry axis). The different symbols denote different
  optical depths between the inner and the outer radius of the
  disk. Legend: $\tau_{\rm ES} = 0.1$ (black circles), $\tau_{\rm ES}
  = 0.4$ (red squares), $\tau_{\rm ES} = 0.7$ (green diamonds),
  $\tau_{\rm ES} = 1.0$ (orange triangles with points up), $\tau_{\rm
  ES} = 1.4$ (blue triangles with points to the left), $\tau_{\rm ES}
  = 3$ (maroon triangles with points to the right), and $\tau_{\rm ES}
  = 5$ (brown crosses)}.
  \label{fig19}
\end{figure}

For a type-1 viewing direction with $i \sim 30\degr$ the maximum
polarization is obtained for values of $\theta_{\rm D}$ in the range of
$20\degr$ -- $30\degr$. It turns out that changing the optical depth
from $\tau \sim 1$ decreases the polarization. To illustrate this, we
plot in Fig.~\ref{fig19} the polarization degree for a flared disk
with $\theta_{\rm D} = 25\degr$ versus the inclination $i$ for various
optical depths.

As in Fig.~\ref{fig18} (top), the polarization curve shows a
discontinuity when the line-of-sight passes the horizon of the
flared disk. However, with larger optical depths, the jump in
polarization becomes shallower because multiple scattering within
the disk has a depolarizing effect. For very low optical depths,
$\tau < 0.1$, the scattering disk is practically invisible, as one
would expect.

The flared disk has somewhat similar polarization properties to
the emitting and scattering disks investigated in
section~\ref{sec:emiscattdisk}. The irradiation pattern is different
but is again axisymmetric. Therefore, at low and moderate optical
depths, the polarization is again mainly defined by photons
traveling in a direction parallel to the disk plane being scattered
at orthogonal scattering angles.  As for emitting and scattering
disks, this produces type-1 polarization. The central funnel of the
flared disk presents a similar geometry to the dusty tori discussed
in section~\ref{sec:torus-HOA}. For flat tori this produces again
type-1 polarization. Because of the funnel, the source is directly
visible at type-1 viewing angles, which has a strong depolarizing
effect. This explains, why the net polarization of flared disks at
type-1 inclinations remains rather small.

\subsubsection{Other geometries}

Finally, we investigate how the geometry of the equatorial
scattering regions influences the polarization for the shapes in
Fig.~\ref{fig17}. For all three shapes the optical depth is defined
horizontally in the equatorial plane between the inner and the outer
radius of the scattering region, and fixed at $\tau_{\rm ES} = 1$. The
inner and outer radii are equal for the three different geometries,
as are the half-opening angles of $\theta_{\rm D} = 30\degr$. The
inclination of the system is assumed to be at the maximal type-1
viewing angle of $\sim 30\degr$.

\begin{figure}
 \vskip 0.5cm \resizebox{8cm}{!}{\includegraphics{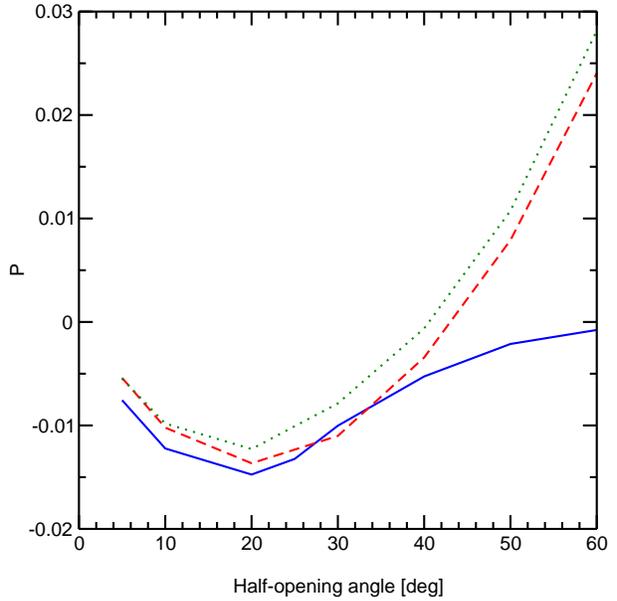}}
 \caption{The effect of the half-opening angle of the equatorial
electron scattering region on polarization for three different
geometries viewed from the maximal type-1 viewing position. Legend:
flared disk (blue solid line), scattering cylindrical disk (dashed red
line), and scattering torus (green dotted line). A negative
polarization denotes polarization parallel to the axis of symmetry.}
 \label{fig20}
\end{figure}

Fig.~\ref{fig20} shows the net polarization versus $\theta_{\rm D}$,
defined for the individual shapes as shown in Fig.~\ref{fig17}.
It can be seen that except for thick flared disks at large
opening angles, there is a similar dependence of $P$ on the
half-opening angle. Changing the geometrical shape of the scattering
disk shifts the overall level of polarization somewhat, but the
dependence on the opening angle largely remains the same. In all
cases the scattering regions produce a polarization vector aligned
with the axis of symmetry at type-1 viewing angles.

We have only considered the possibility of electron scattering close
to the black hole, and not dust scattering, for several reasons. The
main reason is that dust will not survive this close to the central
continuum source. IR reverberation mapping puts the inner edge of
the dusty torus just outside the broad-line region in agreement with
the predicted sublimation temperature (Suganuma et al.
\cite{suganuma06}). If we replace the electrons in the models
discussed in this section with dust it does not give a high enough
polarization (less than around 1\% even for highly-inclined viewing
positions and around 0.1\% at best for more likely near face-on,
type-1 viewing angles) because the dust is forward scattering.
Since there is absorption for dust grains the number of photons
coming out is also lower than in the electron-scattering case.

In summary, the polarization arising from equatorial scattering
regions has the correct orientation of the $\vec E$-vector for
type-1 objects, but the degree of polarization remains moderate at
type-1 viewing angles. This is largely independent of the geometry
chosen. The low polarizations arise partly because even at maximal
type-1 viewing directions the system is still close to being axisymmetric. The
resulting low polarization is further decreased by dilution with the
unpolarized light coming from the central source. Multiple scattering will
also play a role in depolarizing the radiation. We obtained maximum
polarization values for $\tau_{\rm ES} \sim 1$.

\section{Discussion}

\subsection{Polarization from a dusty, optically-thick torus}

From our modeling of uniform-density tori we obtain polarizations of
a few percent from the torus alone for type-2 objects (i.e., without
the need to evoke electron scattering cones -- although such cones are
clearly present in NGC\,1068). The wavelength dependence of
polarization is rather similar for different geometrical shapes of
the inner torus surface.  A flat surface (case 1), a slightly curved
shape (case 2), or a convex shape (case 3) all reveal a nearly flat
polarization spectrum over the optical and UV range for a broad
range of half-opening angles. Thus, polarimetry alone does not give
the geometry of dust distributions in the central regions of AGN.

We have found that the polarization is somewhat higher for a smaller
steeper torus (see Fig. \ref{fig8}) than for a larger more gradual
one. At the same time, the obscuration efficiency at type-2
viewing angles is much higher for gradually curved tori. A real
torus probably has a steep inside to it. This is because the dust
radius is set by dust sublimation and so will be determined
primarily by the inverse square of the distance from the central
source. The inside of the torus is thus probably concave towards the
central source. The source creates a spherical region with a radius
of the order of the sublimation radius of the dust grains, where the
dust composition is likely to be very different to the dust farther
away. Destruction of small dust grains will modify the grain size
distribution and hence the albedo and scattering properties are
likely to change. It will be an interesting future task to model
polarization by scattering off a dusty torus with a more general
geometry and using dust compositions different from the Galactic
composition we have adopted here.

We have constrained the relation between the polarization from
the torus and the opening angle (see Fig. \ref{fig9}). For narrow
opening-angles, the polarization is stronger and has a polarization
vector perpendicular to the symmetry axis. With increasing
half-opening angle the polarization decreases and then gradually
switches to a parallel polarization vector. Recent derivations for
the number ratio of Seyfert-1 to Seyfert-2 galaxies constrain the
half-opening angle of the torus to a range of 38$\degr$ -- 48$\degr$
depending on the sample and the analysis considered (see e.g.,
Tovmassian \cite{tovmassian2001}, Schmitt et al. \cite{schmitt01}).
Thus, for Seyfert galaxies, a realistic dusty torus can only produce
type-2 polarization.

The fraction of type-2 AGNs decreases with luminosity, declining from
80\% when $\log L$ = 42 to 30\% when $\log L$ = 46. (see Ueda et
al. \cite{ueda03}; Hasinger \cite{hasinger04} for recent
counts). Therefore, Lawrence (\cite{lawrence91}) proposed the receding
torus model, in which higher luminosities of AGNs correspond to larger
torus half-opening angles $\theta_0$. This leads to the prediction
that, on average, the polarization of Seyfert-2 galaxies decreases
with luminosity.  The situation will be complicated by possible
additional polar scattering that enhances type-2 polarization,
however, there has to be a minimum amount of polarization coming from
the torus alone and we predict that this decreases with
luminosity. The case is different for higher luminosity quasars
because there is good evidence that they have larger torus
half-opening angles (see e.g. Simpson 2003). We have shown that if the
torus is large and elliptically shaped, type-2 quasars with $\theta_0
> 60\degr$ may show polarization vectors aligned with the symmetry
axis.  Again, this idea only holds if polar scattering is low, which
does not seem to be the case for recently observed of type-2 quasars
(Zakamska et al. \cite{zakamska2005}). But, in any case, the torus has
to contribute to the observed polarization to some extent, and the
amount of its polarization must increase with the torus opening
angle. A thin, disk-like torus was considered for the
high-polarization quasar OI~287 (Rudy \& Schmidt \cite{rudy88}).

A larger opening angle of the torus corresponds to a smaller
covering factor of the central source with dust. Therefore, one
measure of the opening angle is the relative IR flux since it
depends on the dust covering factor. We predict that the degree of
polarization, on average, should be correlated with the relative
strength of the thermal IR flux. So far, P. Smith et al. (\cite{psmith03})
found a correlation between the K-band luminosity and the broad
band optical polarization of a sample of 2MASS QSOs. Their Fig.~4 shows
that QSOs with higher K-band luminosity have also higher polarization
percentages in the optical and the authors point out that this could be
due to more luminous or more extended scattering regions. Polarization
due to scattering by a dusty torus can explain this correlation
for type-2 objects: the higher K-band flux corresponds to a larger
covering factor of the source and thus to a more narrow opening angle
of the torus. Such tori produce higher degrees of type-2
polarization. In the data shown by P. Smith et al. (\cite{psmith03})
most of the high polarization quasars are indeed of the spectral
type 1.5 or 1.8--2. However, there are also a few type-1 objects
that have a high K-band flux. A larger statistical sample of AGN needs
to be examined for correlations between the optical polarization
and the IR flux in order to draw a solid conclusion on this issue.

The lack of wavelength-dependent polarization we find is important
for interpreting spectropolarimetry of type-2 AGNs.
Wavelength-independent polarization is observed in the nuclei of the
Seyfert-2 galaxy NGC~1068 (Antonucci \& Miller, \cite{antonucci85}),
for example. The wavelength independence of the polarization
close to the nucleus of NGC~1068 suggests electron scattering in a
polar mirror (Miller, Goodrich, \& Mathews, \cite{miller91}; Kinney
et al. (\cite{kinney91}) and even enabled a 3D-decomposition of the
mirror in single electron blobs (Kishimoto \cite{kishimoto99}).
However, it is difficult to maintain the ionization of the outflow,
which requires an intrinsically anisotropic ionizing continuum
(Miller, Goodrich, \& Mathews, \cite{miller91}). We suggest that
wavelength-independent polarization also partly arises from the
walls of the torus and combines with the polar scattering. The fact
that dust scattering can mimic wavelength-independent polarization
has been discussed for distant radio galaxies (Dey et al.
\cite{dey96}; Cimatti et al. \cite{cimatti96}), while it was
stressed that a sharp drop of the dust albedo below 2500~\AA~should
lead to a break in the polarization spectrum. In our modeling, such
a feature is barely visible with the Galactic dust composition we
used. However, the grain model we applied is rather simple
in that it only assumes spherical grains. Most dust models of this type
predict a sharp drop of the albedo below $2500$~\AA~while observationally
constrained dust albedos rise toward shorter wavelengths in the Far-UV
(Gordon \cite{gordon04}). The ambiguity between electron scattering and
dust scattering for the wavelength-dependence of polarization can hence
persist even below 2500~\AA.

\subsection{Polarization from polar scattering cones}

We have modeled polar scattering regions considering of both
dust and electrons. As expected, only type-2 polarization is
obtained, and the polarization degree rises towards edge-on viewing
angles. Larger opening angles of the cone lead to lower polarization
values. However, the results obtained here should still be used with
care for the interpretation of real data. It is an implicit
assumption of our modeling (and of previous modeling) that the
medium is homogeneous and extends over the whole volume of the cone.
This does not need to be the case as was shown for NGC~1068 in which
the ionization cones are only party filled with scattering material
(Capetti \cite{capetti95a}). For cones with larger opening angles,
the detailed distribution of the material and the fact that it might
be clumpy plays a role in determining the net polarization and
simulations should be more detailed (Kishimoto \cite{kishimoto96}).
For instance, Miller, Goodrich, \& Mathews (\cite{miller91}) observe
strongly wavelength-dependent polarization from a cloud near the
nucleus. The polarization rises toward shorter wavelengths. This is
frequently explained by dust scattering because the dust scattering
cross section increases toward the blue as well. However, our
computations for scattering off optically-thin dust shown in
Fig.~\ref{fig14} reveal a different behavior. At wavelengths longer
than $\sim 2500$~\AA~the polarization degree seen at type-2 viewing
angles decreases toward the blue because of the wavelength dependent
polarization phase function (Zubko \& Laor \cite{zubko00}). The
reason that the individual cloud in NGC~1068 shows rising
polarization to shorter wavelengths must therefore rely on dilution
by an unpolarized blue starlight spectrum from the host galaxy.

Another important reason why the polar scattering regions
should not be homogeneous is the resulting net polarization at
type-2 viewing angles. If the cones have a low optical depth, the
expected polarization would be very high as we show in
Fig.~\ref{fig13}. This picture can be accurate only for localized
scattering regions such as the individual clouds resolved in
NGC~1068 which have a polarization up to 60\% (see Fig.~4 in Capetti
\cite{capetti95a}). For distant objects, where the polar regions
cannot be resolved, the observed net polarization is lower.
Therefore, as Kishimoto (\cite{kishimoto96}) pointed out,
multiple-scattering in a clumpy medium should reduce the net
polarization.

We have found in our models that the radial dependence of the
scattering material inside the cones is not very important in determining the
net polarization. The important parameter is the optical depth. This is true
if one considers isolated cones with no other scattering mirrors involved. If
one also assumes that there is an obscuring and reflecting torus, the
situation must change. The torus will have a collimating effect, particularly
on the central parts of the cones, so their central density becomes more
important for the net polarization.

Our study of dust scattering in polar cones again underlines
the importance of the optical depth for the wavelength dependence of the
resulting spectra in polarization and total flux. While an
optically-thick obscuring torus produces polarization that is
roughly constant over the optical and UV range, scattering in cones
induces a characteristic wavelength dependence. This dependence
should vary with the dust composition, and future observations and
modeling could constrain the dust composition.

\subsection{Relative importance of toroidal and polar
  scattering regions}

To a first approximation, the net polarization, $P_{net}$, due
to scattering in polar cones and scattering off the torus is given
by the sum of the polarized fluxes $P_{cones} \times F_{cones}$ and
$P_{torus} \times F_{torus}$ coming from the two scattering regions.
This summation neglects radiative coupling between the two
scattering regions, but if we take the polar cone to be optically
thin, it should give a good estimate. By comparing the two single
polarized fluxes, we can constrain the optical depth of the polar
scattering region at which the torus becomes important for the
polarization. We assume a Seyfert-2 galaxy with a torus half-opening
angle of $\theta_0 = 30\degr$ seen at the inclination $i = 63\degr$.
The polarized flux for the given parameters of a large torus can be
derived as shown in section~\ref{sec:torus}. We have used the same
computational method as in section~\ref{sec:cones} for the polar
cones. The equatorial obscuration is realized by limiting the
emission angle to $\theta_0$. We find that a polar double-cone with
the same opening angle, $\theta_0 = 30\degr$, as the torus can have
an optical depth as low as $\tau_{\rm ES} \sim 5 \times 10^{-7}$ to produce a
comparable polarized flux along $i = 63\degr$.

We also give an estimate for the optical depth of the
scattering cone needed in a Seyfert-2 system with half-opening angle
$\theta_0 = 60\degr$ seen at an inclination of $i =
70\degr$. Again modeling the polarized fluxes of both components
separately, we need an optical depth of $\tau_{\rm ES} \sim 5 \times
10^{-5}$ for the scattering cone to produce a comparable polarization
to the torus.

Our modeling thus shows that it is not possible to
independently distinguish between polarization by the torus and
by polar scattering. As soon as there is even very optically thin
polar scattering material, it dominates the net flux {\em and}
polarization. The very efficient obscuration of large tori stresses
the importance of polar scattering regions for the detectability of
type-2 objects, and at the same time it lowers the importance of the
exact polarization signature at very high inclinations. The
polarized flux scattered by a torus at extreme edge-on viewing
angles is so low that it hardly contributes the net polarization at all.

Both types of scattering regions are closely related, however,
because the torus collimates the light entering the polar cones.
Since the albedo of Galactic dust ranges between 40\% and 60\% a
significant fraction of the light scattered in polar cones has
already been scattered by the torus. To obtain the correct net
polarization it is in principle necessary to conduct
multi-scattering simulations involving both scattering regions. This
is especially important when the base of the cone reaches far inside
the funnel of the torus and has sufficient optical depth, and it
remains true, even if the material in the cones is not homogeneous
but organized in localized blobs as seen in NGC~1068 (Kishimoto
\cite{kishimoto99}). Furthermore, the above estimate does not
include any polarization induced by other scattering regions such as
the accretion disk or equatorial material.

In summary, the polarization of both type-1 and type-2 AGNs arises
in a complex way that will depend on the viewing angle. We predict
that the polarization properties will depend on orientation
indicators such as the radio properties, the width of H$\beta$, and
the presence of various types of intrinsic absorption-line systems.

\subsection{Polarization from the accretion disk and from equatorial
  scattering regions }

For type-1 objects, scattering from dust or electrons in the
polar regions of the object only has a small impact on the observed
polarization because the scattering angle is relatively small and
both the covering factor and optical depth are low. Our modeling
shows that polarization induced by a dusty torus is also very low
along type-1 viewing angles. An additional problem is that
scattering cones only produce ``type-2 polarization'' (i.e.,
perpendicular to the symmetry axis), whilst we find that tori can
only give the correct polarization in type-1 quasars for large
opening angles. To understand the polarization properties of
Seyfert-1 galaxies, it is hence necessary to introduce additional
structures to the common AGN scheme.

We have reproduced the earlier result that an emission and
scattering disk can produce type-1 polarization only when it is
relatively optically thin regardless its geometrical thickness. The
accretion disk itself is therefore ruled out as the source of the
type-1 polarization.  However, the disk does give intrinsic
perpendicular polarization that will be coupled successively with
scattering regions such as the torus or scattering cones.

We confirm that the correct type-1 polarization can be produced by
scattering of an equatorial disk has recently been extensively
analyzed by Smith et al. (\cite{smith02}, \cite{smith04}). The
rotating BLR-disk they introduce together with the surrounding
co-planar scattering region reproduces very well the turnover of the
polarization vector observed in the H$\alpha$ and H$\beta$ lines of
many objects (Smith et al. \cite{smith05}). This gives strong
support for the existence of a flattened equatorial scattering
region. Recently, additional support for the idea comes from the
discovery of a short timescale lag of the variations of the
polarization relative to the unpolarized continuum (Shoji, Gaskell,
\& Goosmann \cite{shoji05}; Gaskell, Shoji, \& Goosmann, in
preparation). We find that the optical depth of the equatorial
scattering material that is required to reproduce the observed
polarization degree in type-1 objects is around unity, which justifies
the approach of Smith et al. \cite{smith05}.

An encouraging result we obtain is that the polarization
properties at type-1 viewing angles do not depend very strongly on
the geometry of the equatorial scattering region. This increases the
flexibility in applying the flared disk model. However, according to
Fig.~\ref{fig20} the optical depth plays a significant role in
determining the polarization. Again, it will be important to
consider the polarization of a flared disk in relation to the other
scattering regions and model all constituents of the AGN
consistently.

\section{Conclusions}

We have developed a versatile Monte Carlo code for modeling polarization
produced by scattering in many astrophysical situations. We have shown that it
reproduces previous results well. The code is publicly available and
can be downloaded from the Internet.

Modeling polarization with a uniform-density torus for
different half-opening angles $\theta_0$ leads to polarization
degrees between 0 \% and 20 \%. The orientation of the $\vec
E$-vector is perpendicular to the symmetry axis for $\theta_0 <
53\degr$ and aligned for $\theta_0 > 60\degr$ at all viewing angles
with $i > \theta_0$. For opening angles in the interval $53\degr <
\theta_0 < 60\degr$, both orientations of the polarization vector
appear at type-2 viewing angles. The polarization behavior is not
strongly sensitive to details of the torus geometry (e.g. cylinders
with sharp edges, versus tori with smooth elliptical
cross-sections). The obscuration efficiency is much higher for
a large than for a compact torus.

While the polarized flux of a torus scattered into edge-on
viewing angles is rather low, we expect its polarization properties
to be relevant for the collimated light entering scattering cones.
This is especially important, if the scattering cones have moderate
optical depths at the base. Since the opening angle of the torus and
the cones increase with luminosity, we predict that the net
polarization of type-2 objects depends on luminosity.

The spectral shape of the polarization produced by scattering off of a torus
is nearly wavelength-independent. If a significant part of the the nuclear
scattering in NGC~1068 is produced by dust in the torus then this removes the
difficulty of having to maintain the degree of ionization of the putative
electron cones.

Polar scattering cones produce perpendicular polarization at
all viewing angles. If they contain dust the wavelength-dependence
of the detected radiation is different for face-on and edge-on
viewing angles. If the dusty cones are seen face-on in transmission,
the spectrum is reddened and the polarization is rather low, while
the scattered spectrum at edge-on directions is blue and more highly
polarized.

The parallel polarization seen in type-1 AGNs arises naturally
from electron scattering by optically-thin, equatorial distributions
around the central source. At a given viewing angle, the
polarization of different types of electron scattering disks does
not depend strongly on the geometry, but it is a strong function of
optical depth. A relatively low optical depth of only a few tenths
can produce the observed polarization. On the other hand, optically
thick equatorial material such an accretion disk only produces
type-2 polarization at all viewing angles.

All scattering regions interact, and the overall polarization
properties depend on consistently modeling all components at the
same time. Although the polarized flux by scattering off an
optically-thick torus is weak at type-2 viewing angles, it has an
important impact on the observed polarization as it collimates the
radiation entering other scattering regions. The same is true for
the accretion disk. The polarization properties of the individual
scattering regions all strongly depend on the inclination of the
system. We therefore expect, that the polarization of AGN changes
systematically with orientation parameters. To quantify this
prediction one needs polarization modeling for multi-component
systems. Such modeling will be discussed in a follow-up article
(Goosmann \& Gaskell, in preparation).

\begin{acknowledgements}

We are grateful to the University of Nebraska Computer Science
department for providing access to the University of Nebraska`s {\it
Prairiefire} supercomputer. David Swanson and Mako Furukawa provided
valuable assistance in transferring the code to {\it Prairiefire}.

We wish to thank Ski Antonucci, Makoto Kishimoto, and the 
referee Karl Misselt, for very detailed and constructive comments on
the manuscript.  We also wish to acknowledge helpful discussions
with Mark Bottorff and Shoji Masatoshi on the structure of AGNs, and
with Tim Gay and Paul Finkler on the atomic physics of various
scattering phenomena. RWG thanks the Department of Physics and
Astronomy of the University of Nebraska for their hospitality.

This research has been supported by the Center for Theoretical
Astrophysics in Prague, by the Hans-B\"ockler-Stiftung in
Germany, and by the US National Science Foundation through grant AST
03-07912.

\end{acknowledgements}


\begin{thebibliography}{}

\bibitem[1996]{agol96} Agol, E., \& Blaes, O.\ 1996, \mnras, 282, 965

\bibitem[1969]{angel69} Angel, J. R. P.\ 1969, ApJ, 158, 219

\bibitem[1976]{angel76} Angel, J.~R.~P., Stockman, H.~S., Woolf,
  N.~J., Beaver, E.~A., \& Martin, P.~G.\ 1976, \apjl, 206, L5

\bibitem[1982]{antonucci82} Antonucci, R. R. J.\ 1982, Nature, 299, 605

\bibitem[1983]{antonucci83} Antonucci, R. R. J.\ 1983, Nature, 303, 158

\bibitem[1983]{antonucci84} Antonucci, R.~R.~J.\ 1984, \apj, 278, 499

\bibitem[1985]{antonucci85} Antonucci, R. R. J., \& Miller, J. S.\
 1985, \apj, 297, 621

\bibitem[1993]{antonucci93} Antonucci, R. R. J.\ 1993, ARA\&A, 31, 473

\bibitem[2002]{antonucci02} Antonucci, R. R. J.\ 2002, in
 Astrophysical spectropolarimetry, ed. J. Trujillo-Bueno,
 F. Moreno-Insertis, \& F. Sánchez (Cambridge, UK: Cambridge
 University Press), 151


\bibitem[1996]{blaes96} Blaes, O., \& Agol, E.\ 1996, \apjl, 469, L41

\bibitem[1983]{bohren83} Bohren, C. F., \& Huffman, D. R.\ 1983,  in
 "Absorption and Scattering of Light by Small Particles", Wiley,
 New York

\bibitem[1977]{brown77} Brown, J.~C., \& McLean, I.~S.\ 1977,
  \aap, 57, 141

\bibitem[1995a]{capetti95a} Capetti, A., Axon, D.~J., Macchetto, F., Sparks,
  W.~B., \& Boksenberg, A.\ 1995a, \apj, 446, 155

\bibitem[1995b]{capetti95b} Capetti, A., Macchetto, F., Axon, D.~J., Sparks,
  W.~B., \& Boksenberg, A.\ 1995b, \apjl, 452, L87

\bibitem[1959]{cashwell59} Cashwell, E. D., \& Everett, C. J.\ 1959,
 in "Monte Carlo Methods for Random-Walk Problems", Pergamon Press

\bibitem[1960]{chandrasekhar60} Chandrasekhar S.\ 1960, Radiative
  Transfer, New York, Dover

\bibitem[1996]{cimatti96} Cimatti, A., Dey, A., van Breugel, W., Antonucci,
  R., \& Spinrad, H.\ 1996, \apj, 465, 145

\bibitem[1995]{cohen95} Cohen, M.~H., Ogle, P.~M., Tran, H.~D., Vermeulen,
  R.~C., Miller, J.~S., Goodrich, R.~W., \& Martel, A.~R.\ 1995, \apjl, 448,
  L77

\bibitem[2004]{czerny2004} Czerny, B., Li, J., Loska, Z., \& Szczerba,
  R.\ 2004, \mnras, 348, L54

\bibitem[1996]{dey96} Dey, A., Cimatti, A., van Breugel, W., Antonucci, R., \&
  Spinrad, H.\ 1996, \apj, 465, 157

\bibitem[1985]{dezotti85} De Zotti, G. \& Gaskell C. M.\ 1985, A\&A,
147, 1

\bibitem[1966]{Dibai66}Dibai, E. A. \& Shakhovskoy, N. M.\ 1966,
  Astronomicheskij Tsirkulyar, 375, 1

\bibitem[1984]{draine84} Draine, B. T., \& Lee, H. M.\ 1984, ApJ, 285,
89

\bibitem[2003]{draine03} Draine, B.~T.\ 2003, \apj, 598,
1017


\bibitem[1996]{ferrarese96} Ferrarese, L., Ford, H. C., \& Jaffe,
W.\ 1996, ApJ, 470, 444

\bibitem[1994] {fischer94} Fischer, O., Henning, Th., \& Yorke, H.
W.\ 1994, A\&A, 284, 187

\bibitem[1992] {ford92} Ford, H. C., Caganoff, S., Kriss, G. A.,
Tsvetanov, Z., \& Evans, I. N.\ 1992, BAAS, 24, 818

\bibitem[2004]{gaskell04} Gaskell, C. M., Goosmann, R. W., Antonucci,
 R. R. J., \& Whysong, D.\ 2004, ApJ, 616, 147

\bibitem[2006]{gaskell06} Gaskell C. M., \& Benker A. J.\ 2006, submitted

\bibitem[2004]{glass04} Glass, I. S.\ 2004, MNRAS, 350, 1049

\bibitem[1994] {goodrich94} Goodrich, R. W., \& Miller, J. S.\ 1994, ApJ, 434,
  82

\bibitem[1995]{goodrich95} Goodrich, R.~W., \& Miller, J.~S.\ 1995, \apjl,
  448, L73


\bibitem[2004]{gordon04} Gordon, K.~D.\ 2004, ASP Conf.~Ser.~309:
Astrophysics of Dust, 309, 77


\bibitem[2004]{hasinger04} Hasinger, G.\ 2004, Nuclear Physics B Proceedings
  Supplements, 132, 86

\bibitem[1997]{heisler97} Heisler, C.~A., Lumsden, S.~L., \& Bailey, J.~A.\
  1997, \nat, 385, 700



\bibitem[1993]{hines93} Hines, D. C., \& Wills, B. J.\ 1993, \apj, 415, 82

\bibitem[1995]{hines95} Hines, D.~C., \& Wills, B.~J.\ 1995, \apjl, 448, L69


\bibitem[1995]{kartje95} Kartje, J. F.\ 1995, \apj, 452, 565

\bibitem[1994]{kay94} Kay, L.~E.\ 1994, \apj, 430, 196

\bibitem[1980]{keel80} Keel, W.~C.\ 1980, \aj, 85, 198

\bibitem[1991]{kinney91} Kinney, A.~L., Antonucci, R.~R.~J., Ward, M.~J.,
  Wilson, A.~S., \& Whittle, M.\ 1991, \apj, 377, 100

\bibitem[1996]{kishimoto96} Kishimoto, M.\ 1996, \apj, 468, 606

\bibitem[1999]{kishimoto99} Kishimoto, M.\ 1999, \apj, 518, 676

\bibitem[2001]{kishimoto01} Kishimoto, M., Antonucci, R., Cimatti, A.,
  Hurt, T., Dey, A., van Breugel, W., \& Spinrad, H.\ 2001, \apj, 547, 667

\bibitem[1999]{komossa99} Komossa, S.\ 1999, ISAS Report, p.~149-160,
  T.~Takahashi, H.~Inoue (eds), 149

\bibitem[1982]{lawrence82} Lawrence, A., \& Elvis, M.\ 1982, ApJ, 256,
410

\bibitem[1991]{lawrence91} Lawrence, A.\ 1991, \mnras, 252, 586

\bibitem[1998]{martel98} Martel, A.\ 1998, \apj, 508, 657

\bibitem[1977]{mathis77} Mathis, J. S., Rumpl, W., \& Nordsieck,
 K. H.\ 1977, \apj, 217, 425

\bibitem[1982]{mezger82} Mezger, P.~G., Mathis, J.~S., \& Panagia, N.\
  1982, \aap, 105, 372

\bibitem[1990]{miller1990} Miller, J.~S., \& Goodrich, R.~W.\ 1990, \apj, 355,
  456

\bibitem[1991]{miller91} Miller, J. S., Goodrich, R. W, \&
Mathews, W. G.\ 1991, ApJ, 378, 47


\bibitem[1999]{ogle99} Ogle, P.~M., Cohen, M.~H., Miller,
  J.~S., Tran, H.~D., Goodrich, R.~W., \& Martel, A.~R.\ 1999, \apjs, 125, 1


\bibitem[1997]{packham97} Packham, C., Young, S., Hough, J.~H., Axon, D.~J.,
  \& Bailey, J.~A.\ 1997, \mnras, 288, 375

\bibitem[1992]{pier92} Pier, E.~A.~\& Krolik, J.~H.\ 1992, \apj, 401, 99

\bibitem[1977]{rowan-robinson77} Rowan-Robinson, M.\ 1977, ApJ, 213, 635

\bibitem[1988]{rudy88} Rudy, R.~J., \& Schmidt, G.~D.\ 1988, \apj, 331, 325

\bibitem[2001]{schmitt01} Schmitt, H.~R.,
  Antonucci, R.~R.~J., Ulvestad,
  J.~S., Kinney, A.~L., Clarke,
  C.~J., \& Pringle, J.~E.\ 2001,
  \apj, 555, 663

\bibitem[1973]{shakura73} Shakura, N.~I., \& Sunyaev, R.~A.\ 1973, \aap, 24,
  337

\bibitem[2005]{shoji05} Shoji, M., Gaskell, C. M., \& Goosmann, R.
W.\ 2005, \baas, 37, 1420

\bibitem[2005]{simpson05} Simpson, C.\ 2005, \mnras, 360, 565

\bibitem[2002]{smith02} Smith, J. E., Young, S., Robinson, A.,
  Corbett, E. A., Giannuzzo, M. E., Axon, D. J., \& Hough\ 2002,
  \mnras, 335, 773

\bibitem[2004]{smith04} Smith, J. E., Robinson, A., Alexander, D. M.,
  Young, S., Axon, D. J., Corbett, Elizabeth A.\ 2004, MNRAS, 350, 140

\bibitem[2005]{smith05} Smith, J. E., Robinson, A., Young, S.,
Axon, D. J., Corbett, Elizabeth A.\ 2005, MNRAS, 359, 846

\bibitem[2003]{psmith03} Smith, P.~S., Schmidt, G.~D., Hines, D.~C.,
\& Foltz, C.~B.\ 2003, \apj, 593, 676

\bibitem[1979]{stockman79} Stockman, H.~S.,
Angel, J.~R.~P., \& Miley, G.~K.\ 1979, \apjl, 227, L55

\bibitem[2006]{suganuma06} Suganuma, M., et al.\
2006, \apj, 639, 46

\bibitem[1985]{sunyaev85} Sunyaev, R.~A., \& Titarchuk, L.~G.\ 1985,
  \aap, 143, 374

\bibitem[2001]{tovmassian2001} Tovmassian, H.~M.\ 2001,
Astronomische Nachrichten, 322, 87

\bibitem[1992]{tran92} Tran, H.~D., Miller, J.~S., \& Kay, L.~E.\ 1992, \apj,
  397, 452

\bibitem[1999]{tran99} Tran, H. D., Brotherton, M. S., Stanford,
 S. A., van Breugel, W., Dey, A., Stern, D., \& Antonucci,
 R. R. J.\ 1999, \apj, 516, 85

\bibitem[2001]{tran01} Tran, H.~D.\ 2001, \apjl, 554, L19

\bibitem[2003]{ueda03} Ueda, Y., Akiyama, M., Ohta, K., \& Miyaji, T.\ 2003,
  \apj, 598, 886

\bibitem[1966]{Walker66} Walker, M. F.\ 1966, \apj71, 184

\bibitem[2006]{wang06} Wang, H.-Y., Wang, T.-G., \& Wang, J.-X.\
accepted by ApJS, astro-ph/0609469

\bibitem[2003]{watanabe03} Watanabe, M., Nagata, T., Sato, S., Nakaya,
H., \& Hough, J.~H.\ 2003, \apj, 591, 714




\bibitem[1999]{wolf99} Wolf, S., \& Henning, T.\ 1999, A\&A, 341, 675


\bibitem[2003]{wolf03} Wolf, S.\ 2003, \apj, 582, 859

\bibitem[1995]{young95} Young, S.,  Hough, J. H., Axon, D. J., Bailey,
J. A., \& Ward, M. J.1995 \mnras, 272, 513

\bibitem[1996]{young96a} Young, S., Hough, J. H., Efstathiou, A.,
  Wills, B. J., Bailey, J. A., Ward, M. J., Axon, D. J.\ 1996a,
  \mnras, 281, 1206

\bibitem[1996]{young96b} Young, S., Hough, J.~H.,
Efstathiou, A., Wills, B.~J., Axon, D.~J., Bailey, J.~A., \& Ward, M.~J.\
1996b, \mnras, 279, L72

\bibitem[1999]{young99} Young, S., Corbett, E. A., Giannuzzo, M. E., Hough,
  J. H., Robinson, A., Bailey, J. A., \& Axon, D. J.\ 1999, \mnras, 303, 227

\bibitem[2000]{young00} Young, S.\ 2000 \mnras, 312, 567

\bibitem[2000]{zubko00} Zubko, V. G. \& Laor, A.\ 2000, ApJS, 128, 245

\bibitem[2005]{zakamska2005} Zakamska, N.~L., et
al.\ 2005, \aj, 129, 1212


\end{thebibliography}
\end{document}